\documentclass[letterpaper]{article}

\usepackage[T1]{fontenc}

\usepackage{geometry}
\usepackage{setspace}

\usepackage[style = chem-acs, doi=true, maxbibnames=75, articletitle=true]{biblatex}
\usepackage{graphicx}
\usepackage{float}
\newfloat{scheme}{htbp}{los}
\floatname{scheme}{Scheme}
\floatname{chart}{Chart}
\newfloat{graph}{htbp}{loh}

\usepackage{chemformula} 
\usepackage[version = 4]{mhchem} 

\usepackage[colorinlistoftodos]{todonotes}
\usepackage[title]{appendix}
\usepackage{multicol}
\usepackage{multirow}                      
\usepackage{hhline}
\usepackage{tabularx}

\usepackage{fontawesome5}

\usepackage{authblk}
\author[1]{Daniel Beer}
\author[2]{Jonas Weiser}
\author[3]{Tom Gabler}
\author[3]{Kirsten Zeitler}
\author[1]{Carsten Deibel}
\author[2]{Christian Wiebeler*}
\affil[1]{Institut für Physik, Technische Universität Chemnitz, 09126 Chemnitz, Germany}
\affil[2]{Institut für Physik \& Center for Advanced Analytics and Predictive Sciences, Universität Augsburg, 86159 Augsburg, Germany}
\affil[3]{Institut für Organische Chemie, Universität Leipzig, 04103 Leipzig, Germany}

\title{Beyond the Static Approximation: Assessing the Impact of Conformational and Kinetic Broadening on the Description of TADF Emitters}
\date{*Email: christian.wiebeler@uni-a.de}

\begin{document}

\begin{refsection}

\maketitle

\begin{abstract}
  \noindent Thermally activated delayed fluorescence (TADF) is a promising route towards high-efficiency, metal-free organic light-emitting diodes (OLEDs). However, the characterization of TADF kinetics in solid-state thin films is often complicated by pronounced multiexponential photoluminescence decays that prevent standard biexponential modeling. In this work, we introduce the 'Gamma-Fit' method, a streamlined analytical framework based on the gamma distribution that accounts for the continuous distribution of decay rates inherent in disordered molecular ensembles. By treating the decay as a result of conformational and kinetic heterogeneity, we accurately extract kinetic parameters for the benchmark emitters 4CzIPN and 5CzBN, as well as a series of novel diphenylamine (DPA)-based systems. Our results reveal that accounting for the local environment in thin films remains an important part in determining OLED efficiency. The experimental findings are complemented by a semiclassical Marcus-like computational approach. We evaluate the reliability of this conventional single-conformation rate calculation method and highlight the presence of conformational ensembles and multiple RISC-active triplet states as important factors for accurately describing the transition kinetics.

\end{abstract}

\section*{Keywords}

\noindent Photoluminescence, Thermally Activated Delayed Flourescence (TADF), Excited States, Kinetic Modeling, Quantum Chemical Calculations, Organic Light-Emitting Diodes (OLEDs), Donor--Acceptor Molecules



\section*{Introduction}

The search for metal-free organic light-emitting diodes (OLEDs) with $100\%$ internal quantum efficiency has positioned thermally activated delayed fluorescence (TADF) as a cornerstone of modern molecular electronics. \cite{Uoyama2012, DosSantos2024} Central to its efficiency is the reverse intersystem crossing (RISC) process, which converts non-radiative triplet excitons into radiative singlets. The fundamental photophysics of TADF is well understood in dilute solutions. However, especially kinetic aspects are more complex in solid-state thin films. \cite{Haase2018, Kelly2022, Serevicius2023} This challenges analytical and computational models.

The efficiency of RISC is governed by the delicate balance of the donor-acceptor (D--A) architecture. High D and/or A strength enhances the charge-transfer (CT) character, minimizing the singlet--triplet energy gap ($\Delta E_\mathrm{ST}$) to facilitate efficient delayed fluorescence. \cite{Im2017, Uoyama2012} However, this process is sensitive to the dihedral angle between the D and A units. \cite{Shi2022, Weissenseel2019, souza2025dynamics} While a nearly orthogonal D--A arrangement minimizes $\Delta E_\mathrm{ST}$, it simultaneously reduces the orbital overlap required for efficient transitions, creating a situation where subtle conformational shifts can drastically alter the RISC rate. \cite{Huang2018, Stachelek2019} Even small thermal fluctuations at room temperature can potentially sample a wide range of D--A orientations, as the molecular geometry probed in experiments is not fixed in conformational space but representative of a distribution across the potential energy surface (PES). \cite{de2024tadf}

This sensitivity is exemplified by the benchmark molecule 2,4,5,6-Tet\-ra\-kis(9\textit{H}-car\-ba\-zol-9-yl)iso\-phthalo\-nitrile (4CzIPN) and its halogenated derivatives. \cite{Uoyama2012, Niwa2014, Ishimatsu2013, Streiter2020} For instance, chlorination has been shown to decrease $\Delta E_\mathrm{ST}$ and accelerate intersystem crossing (ISC), yet it often compromises the photoluminescence quantum yield (PLQY) through increased non-radiative pathways. \cite{Streiter2020} Similarly, bromination in systems such as 3-PXZ-XO (3-(10H-phenoxazin-10-yl)-9\textit{H}-xan\-then-9-one) increases both ISC and RISC rates, yet unexpectedly yields an opposite trend in ACRXTN (3-(9,9-di\-methyl\-ac\-ri\-din-10(9H)-yl)-9\textit{H}-xan\-then-9-one) due to changes in the CT state. \cite{Aizawa2020} These diverging results underscore the fact that heavy-atom effects for D--A type TADF molecules appear to be secondary to the primary influence of molecular conformation and electronic environment.

Conventionally, the computational determination of ISC and RISC rates has relied on a simplified framework derived from Fermi’s Golden Rule, typically expressed in a form analogous to Marcus Theory. \cite{marcus1993electron, bredas2004charge, Olivier2017} In this approach, the transition rate is primarily governed by the energy splitting $\Delta E_{ST}$ between the lowest excited singlet ($S_\mathrm{1}$) and triplet ($T_\mathrm{1}$) states and the corresponding spin--orbit coupling (SOC) matrix element. This methodology assumes that the transition occurs exclusively between the equilibrium minima of these two states, with electronic parameters calculated at the respective optimized geometries. 

Despite its utility, this approach suffers from several limitations that can lead to deviations from experimental observations. First, the reliance on a two-state model ($S_\mathrm{1}$ and $T_\mathrm{1}$) does not take the relevance of higher-lying triplet states ($T_\mathrm{n}$) into account. \cite{Aizawa2020, gao2026unveiling} These states can play a decisive role in the RISC process by providing mediated pathways or spin--vibronic channels that significantly enhance the rate beyond what $\Delta E_{ST}$ alone would suggest. \cite{gibson2017nonadiabatic, kim2019spin} Furthermore, the assumption of static, optimized geometries fails to account for the breakdown of the Condon approximation. In many high-performance emitters, the spin-orbit coupling is not a constant but is also sensitive to the D--A torsional angles. \cite{kaminski2024balancing} This can lead to an inaccurate description of the complex kinetic landscape, especially in disordered thin films.

In thin films, molecules are "frozen" into a vast ensemble of conformations dictated by the local morphology and matrix interactions. Unlike the free rotation possible in solution, solid-state environments lock molecules into a distribution of D--A angles and $\Delta E_\mathrm{ST}$ values. \cite{Stavrou2020, Chen2023} This energetic and structural disorder manifests as a power-law behavior observed in time-resolved photoluminescence (trPL) decays. \cite{Haase2018} While sophisticated tools such as the inverse Laplace transformation can extract these rate distributions, their mathematical complexity often hinders widespread adoption. \cite{Kelly2022, Streiter2020}

To address this issue, we introduce a streamlined 'Gamma-Fit' method designed to capture the multiexponential nature of thin-film TADF decays with a small number of free parameters and reduced complexity compared to the inverse Laplace transforms.
By treating the decay as a distribution of states rather than a discrete sum of exponentials, we accurately model the transition from the near-ideal biexponential behavior of solutions to the complex power-law kinetics of thin films. We demonstrate the versatility of this approach using the benchmark emitters 4CzIPN and 2,3,4,5,6-Penta\-(9\textit{H}-car\-ba\-zol-9-yl)benzo\-nitrile (5CzBN), alongside a series of di\-phenyl\-amine (DPA)-based systems, providing a robust framework for facile extraction of kinetic parameters in disordered organic solids. Furthermore, we evaluate the predictive accuracy of the conventional computational framework for RISC rate calculation, specifically addressing the influence of conformational ensembles and the contributions of multiple RISC-active triplet states.

\begin{figure}[H]
    \centering
    \includegraphics[width=8cm]{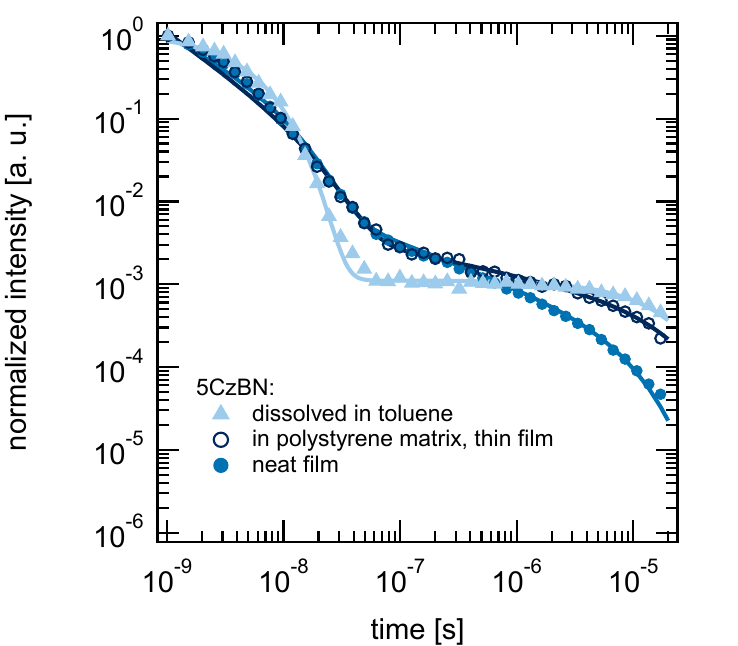}
    \caption{Transient photoluminescence data for 5CzBN dissolved in a toluene solution (triangles), embedded in a polystyrene matrix (open circles) and in a neat film (closed circles). The data were fitted using the 'Gamma-Fit' (solid lines).}
    \label{fig:5Cz_sol_film}
\end{figure} 

\section*{Methods}\label{sec2}
Figure \ref{fig:5Cz_sol_film} shows three exponential excited state decays of 5CzBN in a toluene solution, in a polystyrene matrix, and as a neat film. The TADF decay in toluene solution can be described by a biexponential decay resulting in a single decay rate for both prompt and delayed fluorescence. The thin film measurements in neat film as well as in polystyrene matrix reveal a power-law decay component characteristic of a broad distribution of decay rates. These resulting decays can be described using a multiexponential fit or, more generally, an inverse Laplace transform. \cite{Haase2018, Streiter2020, Kelly2022} However, both methods rely on the summation of $n$ discrete exponential components. To provide a more physically intuitive and continuous representation that circumvents this summation and accurately represents the multiexponential decays, we employ the gamma distribution, a generalization of the exponential distribution that inherently includes a power-law component and an exponentially decaying component. The gamma distribution is defined as:

\begin{equation}
    f(t) = \frac{k^r}{\Gamma(r)} \cdot t^{r-1} \cdot \exp{(-k\cdot t)} \ ,
    \label{eq_gamma_allg}
\end{equation}
\\
where $k$ is the rate parameter, $r$ is the shape parameter, $t$ is time, and $\Gamma(r)$ is the gamma function. \cite{Krishnamoorthy2006}

Physically, the use of a gamma distribution is motivated by the inherent structural and energetic disorder within neat films. Different conformers and varying environmental effects result in a non-uniform distribution of states that is more accurately captured by a tailed distribution than by discrete states. In this framework, $k$ represents the decay rate of the slowest component (dominating the exponential tail), while $r$ characterizes the power-law behavior. For $r=1$, the distribution simplifies to a delta function of rates, resulting in a monoexponential decay. For $r < 1$, the distribution accounts for a broader range of rate parameters, giving rise to the observed power-law component.

To describe the entire normalized TADF decay, we utilize a superposition of two gamma distributions: A temperature-independent distribution for prompt fluorescence (PF) and a temperature-dependent distribution for thermally activated delayed fluorescence (DF): 
\begin{equation}
    \begin{split}
        f_\Gamma(t, T) =&~ \frac{k_\mathrm{PF}^{r_\mathrm{PF}}}{\Gamma(r_\mathrm{PF})} \cdot t^{r_\mathrm{PF}-1} \cdot \exp{(-k_\mathrm{PF}\cdot t)} \\
        &+ A_\mathrm{DF}(T) \cdot \frac{k_\mathrm{DF}(T)^{r_\mathrm{DF}(T)}}{\Gamma(r_\mathrm{DF}(T))} \\&\cdot t^{r_\mathrm{DF}(T)-1} 
        \cdot \exp{(-k_\mathrm{DF}(T)\cdot t)}
    \label{eq_Gamma_TADF}
    \end{split}
\end{equation}
\\ 
Here, $k_\mathrm{PF}$ and $k_\mathrm{DF}$ are the rate constants, and $r_\mathrm{PF}$ and $r_\mathrm{DF}$ are the shape parameters for the prompt and delayed components, respectively. The scaling factor $A_\mathrm{DF}(T)$ adjusts the relative contribution of the delayed fluorescence so that $A_\mathrm{PF} = 1$ for the prompt fluorescence is independent of temperature. To describe the normalized TADF decays, $f_\Gamma(t, T)$ is normalized to $1$ for $t$ approaching $0$. The temperature dependence of these parameters is described in more detail in Section S4 of the Supporting Information.

Using global fits of the temperature-dependent decays, we extracted TADF kinetic parameters following a modified approach based on the work of Tsuchiya et al. \cite{Tsuchiya2021} We assume that the prompt fluorescence lifetime $k_\mathrm{p}$ corresponds to the mean value of its respective gamma distribution. The prompt fluorescence efficiency $\Phi_\mathrm{PF}$ is then determined by integration of the gamma distribution. The delayed fluorescence parameters $k_\mathrm{d}$ and $\Phi_\mathrm{DF}$ are determined the same way. Intersystem crossing and reverse intersystem crossing rates are subsequently calculated as: 
\begin{align}
    k_\mathrm{ISC} &= k_\mathrm{p}\cdot\frac{\Phi_\mathrm{DF}}{\Phi_\mathrm{PLQY}}-k_\mathrm{d}\cdot\frac{\Phi_\mathrm{DF}}{\Phi_\mathrm{PF}} \label{eq_k_isc}\\
    k_\mathrm{RISC} &= k_\mathrm{d}\cdot \frac{\Phi_\mathrm{PLQY}}{\Phi_\mathrm{PF}} \label{eq_k_risc}
\end{align}
with the PLQY $\Phi_\mathrm{PLQY}$. Phosphorescence has been neglected in all our calculations. For more details on our approach, see Section S5 of the Supporting Information.
We further determine the activation energy $E_\mathrm{A}$ for the RISC process using an Arrhenius relationship:
\begin{equation}
    k_\mathrm{RISC} = k_\mathrm{A}\cdot\exp \left(-\frac{E_\mathrm{A}}{k_\mathrm{B} T}\right) \ ,
    \label{eq:k_risc}
\end{equation}
\\
with the Boltzmann constant $k_\mathrm{B}$, the temperature $T$ and the transition rate $k_\mathrm{A}$. The corresponding Arrhenius plots are shown in Figure S6. \cite{Uoyama2012, Dias2017} 
Eq (\ref{eq:k_risc}) can be cast in a semiclassical Marcus-like expression: \cite{marcus1993electron, bredas2004charge, Olivier2017}
\begin{equation}
    \begin{split}
        k_\mathrm{RISC} = &\frac{2\pi}{\hbar} \mid H_\mathrm{SO}\mid^2 \left(4\pi\lambda k_\mathrm{B} T\right)^{-\frac{1}{2}}\\
        &\cdot\exp \left(-\frac{E_\mathrm{A}}{k_\mathrm{B} T}\right) \ ,
        \label{eq:H_SO}        
    \end{split}
\end{equation}
\\
where $\lambda$ is the reorganization energy and $H_\mathrm{SO}$ is the SOC matrix element. The SOC can be determined using eq (\ref{eq:H_SO}) and the reorganization energy obtained from the excitation and emission spectra. Additionally, the energy gap between the singlet and triplet states $\Delta E_\mathrm{ST}$ is related to the activation energy by: 
\begin{equation}
    E_\mathrm{A} = \frac{(\Delta E_\mathrm{ST} + \lambda)^2}{4\lambda}
    \label{eq:E_A}
\end{equation}

Finally, to validate these experimental findings, we performed quantum chemical calculations at the time-dependent density functional theory (TD--DFT) level and used the Marcus-like approach introduced in eq (\ref{eq:H_SO}) to calculate RISC and ISC rates for all molecules. Details regarding the preparation of minimum energy geometries, the extraction of the relevant electronic properties (SOC, reorganization energy, singlet--triplet gap), and the subsequent rate calculations can be found in the Experimental and Computational Details Section.
\begin{figure}[H]
    \centering    
    \includegraphics[width=\textwidth]{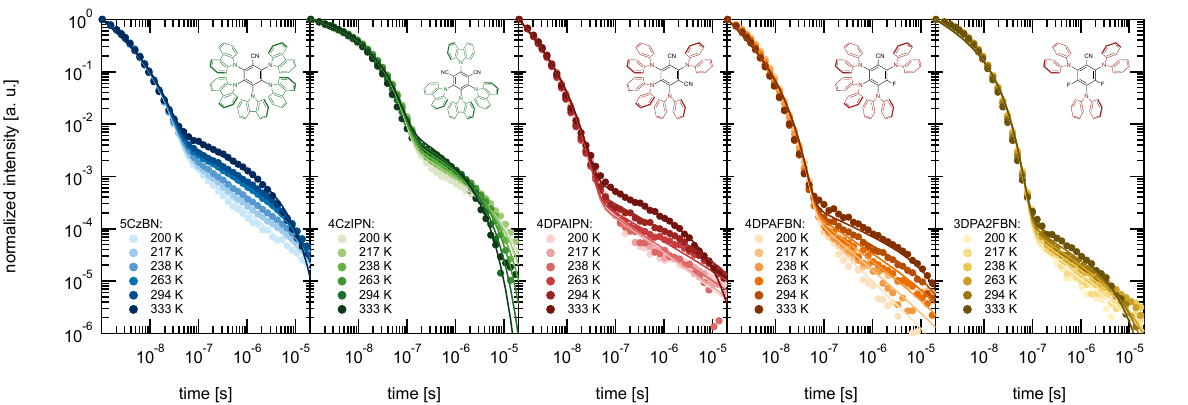}
    \caption{Temperature-dependent transient photoluminescence data (dots) of 5CzBN, 4CzIPN, 4DPAIPN, 4DPAFBN, and 3DPA2FBN fitted using the 'Gamma-Fit' method (lines). All measurements show an increase in delayed fluorescence with higher temperatures. Carbazole-based compounds exhibit higher delayed fluorescence intensities compared to diphenylamine-based compounds. Carbazole and diphenylamine donors are depicted in green and red, respectively.}    
    \label{fig:all_neben}
\end{figure}

\section*{Results and discussion}

The novel method outlined in the preceding section was applied to a variety of TADF emitters. In addition to 5CzBN and 4CzIPN, the molecules 2,4,5,6-Tet\-rakis(di\-phenyl\-amino)iso\-phthalo\-nitrile (4DPAIPN), 2,3,4,6-Tet\-ra(di\-phenyl\-amino)-5-fluoro\-benzo\-nitrile (4DPAFBN) and 2,4,6-Tris(di\-phenyl\-amino)-3,5-di\-fluoro\-benzo\-nitrile (3DPA2FBN) were investigated.  All molecules were synthesized according to previous work. \cite{Speckmeier2018, Garreau2019, Chernowsky2021} Further details are given in Sections S1 to S3 of the Supporting Information. The molecular structures of the five molecules are shown in Figure \ref{fig:all_neben} next to their corresponding transient trPL decays. A comparison of the five materials shows extended prompt fluorescence  for the carbazole (Cz)-based compounds, while the PL intensity of the DPA-based compounds drops by up to four orders of magnitude within the first $100\,\mathrm{ns}$. At around $100\,\mathrm{ns}$, delayed fluorescence becomes dominant for all molecules, with the DPA-based emitters exhibiting a longer comparative lifetime. Additionally, the power-law decay is more pronounced in these compounds, which indicates a broader distribution of decay rates (see below). All decays were fitted using the 'Gamma-Fit' method outlined in eq (\ref{eq_Gamma_TADF}), resulting in good agreement of the fits with the raw data. The characteristic parameters for TADF decays were then extracted from these fits as described in the previous section (for details, see Table S1).

For 5CzBN and 4CzIPN, the radiative singlet recombination rates were determined to be $108\cdot 10^6\,\mathrm{s^{-1}}$ and $28.0\cdot 10^6\,\mathrm{s^{-1}}$, respectively. The PLQY of the neat thin films were determined to be approximately $75\,\%$ while the non-radiative recombination rate ($k_\mathrm{nr}^\mathrm{S}$) was determined to be approximately one third of the radiative singlet recombination rate ($k_\mathrm{r}^\mathrm{S}$) for both systems. The extended singlet lifetime of 5CzBN allows for more singlets to be converted into triplets. This results in an increased $\Phi_\mathrm{DF}/\Phi_\mathrm{PF}$ fraction, rising from $0.45$ in 4CzIPN to $1.1$ in 5CzBN. The singlet-to-triplet conversion is further enhanced by a larger ISC rate in 5CzBN. However, the RISC rate is significantly smaller due to a wide singlet--triplet gap of $\Delta E_\mathrm{ST}=120\,\mathrm{meV}$ ($\Delta E_\mathrm{ST}=58\,\mathrm{meV}$ for 4CzIPN). Overall, these values are consistent with those reported in the existing literature ($110$--$170\,\mathrm{meV}$ for 5CzBN \cite{Hosokai2018, Cho2020, Huang2024, Noda2018} and $40$--$85\,\mathrm{meV}$ for 4CzIPN \cite{Uoyama2012, Streiter2020, Olivier2017}, respectively), which confirms the suitability of our approach.

Next, to clarify how different donor units modulate radiative and non-radiative processes to the ground state, we contrasted the properties of 4CzIPN with those of 4DPAIPN. As shown in Figure \ref{fig:all_neben}, the PF decreases more rapidly in 4DPAIPN compared to 4CzIPN. We can explain this with a fast radiative recombination rate of $65.9\cdot10^6 \,\mathrm{s^{-1}}$ for 4DPAIPN, which is increased by a factor of three compared to 4CzIPN, as well as an even faster non-radiative recombination rate of $88.9\cdot10^6 \,\mathrm{s^{-1}}$. The stronger influence of non-radiative recombination in 4DPAIPN is also reflected in the PLQY, which decreases to $42.6\,\%$. The enhancement of non-radiative recombination can be attributed to the flexible DPA donor units, which possess additional low-frequency vibrational modes due to the presence of unconstrained aromatic rings. This results in a greater number of non-radiative decay channels, leading to an increase in singlet recombination. This is also reflected in the $\Phi_\mathrm{DF}/\Phi_\mathrm{PF}$ ratio, which is smaller than $0.01$ due to a low ISC rate of $1.21\cdot10^6 \,\mathrm{s^{-1}}$. This comparatively low rate can be attributed to unfavourable alignment of singlet and triplet states exemplified by a high $\Delta E_\mathrm{ST}$ value of $146\,\mathrm{meV}$. A similarly low RISC rate of $24.0\cdot10^4 \,\mathrm{s^{-1}}$ rules out 4DPAIPN as an effective TADF emitter.

In addition, we evaluated the influence of fluorination on DPA-based emitters by comparing the properties of 4DPAIPN, 4DPAFBN, and 3DPA2FBN. While both fluorine-containing molecules exhibit radiative recombination similar to that of 4DPAIPN, the non-radiative recombination rate of 4DPAFBN ($438\cdot10^6 \,\mathrm{s^{-1}}$)  is approximately five times faster than that of 3DPA2FBN and 4DPAIPN ($90.8\cdot10^6 \,\mathrm{s^{-1}}$ and $88.9\cdot10^6 \,\mathrm{s^{-1}}$, respectively). This results in a low PLQY of $12.6\,\%$  for 4DPAFBN, compared to $42.6\,\%$ and $40.4\,\%$ for 4DPAIPN and 3DPA2FBN, respectively. Additionally, the contribution of delayed fluorescence is low for both 4DPAFBN and 3DPA2FBN due to rapid singlet recombination ($\Phi_\mathrm{DF}/\Phi_\mathrm{PF} < 0.1$) analogous to 4DPAIPN. However, the ISC rate of 4DPAFBN is more than ten times higher than those of 3DPA2FBN and 4DPAIPN. Conversely, RISC is more effective in 3DPA2FBN and 4DPAIPN than in 4DPAFBN due to narrower singlet--triplet gaps of $104\,\mathrm{meV}$ and $146\,\mathrm{meV}$, respectively. The effect of halogenation of cyanoarene systems on $\Delta E_\mathrm{ST}$ and RISC rate has also been demonstrated in other publications. \cite{Streiter2020, Aizawa2020}.

\begin{figure}[H]
    \centering
    \includegraphics[width=8cm]{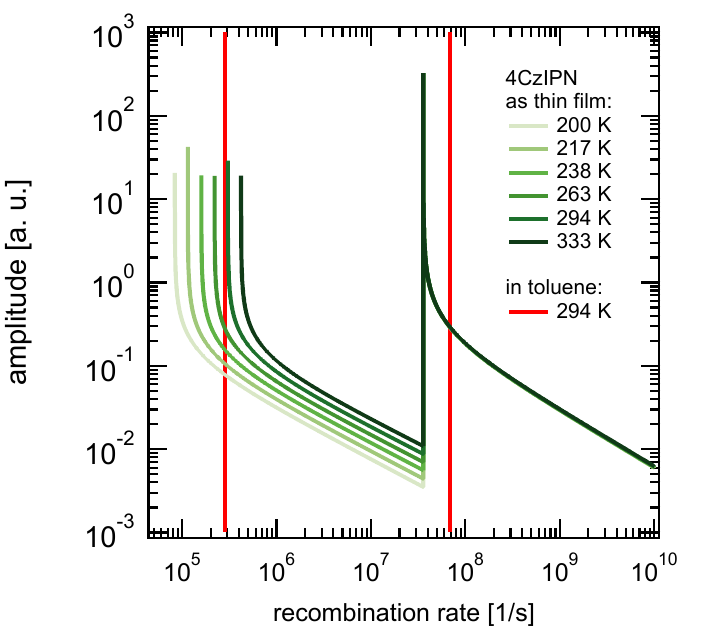}
    \caption{Distribution of decay rates for 4CzIPN at temperatures between $200\,\mathrm{K}$ and $333\,\mathrm{K}$ determined by the 'Gamma-Fit' method. In addition to the dominant peaks, the power-law decay of the data results in contributions from faster rates that decrease with the form factor $r$. A biexponential fit could be assumed for measurements in toluene at room temperature, which is shown by two delta peaks (red).}
    \label{fig:rate_distributrion_4CzIPN_temp}
\end{figure}

Now we will focus on the distribution of decay rates extracted from the 'Gamma-Fit' for 4CzIPN. To assess the distribution of decay rates, we applied an inverse Laplace transformation to eq (\ref{eq_Gamma_TADF}), which yields the decay rate distribution for different temperatures (see Figure \ref{fig:rate_distributrion_4CzIPN_temp}). For 4CzIPN, there are dominant contributions of prompt fluorescence at $3.59\cdot 10^7\,\mathrm{s^{-1}}$ and delayed fluorescence between $8.38\cdot 10^4\,\mathrm{s^{-1}}$ and $4.22\cdot 10^5\,\mathrm{s^{-1}}$, respectively. These dominant contributions describe the exponentially decreasing end of the trPL data. In addition to these main peaks, faster rates with $k^{-r}$-decreasing amplitudes contribute to the TADF decay. This means that the TADF decays are predominantly influenced by the exponentially decaying edge, with contributions from faster decay processes affecting the decay shape. This form of decay allows for the analysis of the decay rate distribution, which becomes broader as $r$ decreases. 
For $r \rightarrow 1$, we obtain a delta distribution for the corresponding decay rate, equating to a monoexponential decay. This limiting case can be assumed for an ideal TADF decay, such as 4CzIPN in toluene solution at room temperature, which is represented by two $\delta$ peaks (red) in Figure \ref{fig:rate_distributrion_4CzIPN_temp}. 
These findings are consistent with results from related studies in the literature: De Thieulloy et al.\ demonstrated that a broad distribution of TADF parameters arises in molecular ensembles, whereas delta peaks are expected for the equilibrium geometry. \cite{Thieulloy2025} In a similar way, Kelly et al.\ were able to show that the distribution of RISC rates in solutions is narrower than in thin films. \cite{Kelly2022} In contrast, Streiter et al. applied an inverse Laplace transform using a few discrete decay rates to fit the multiexponential decays. \cite{Streiter2020} However, the distribution of rates in the literature is more symmetrical than in our model. \cite{Kelly2022, Thieulloy2025, Qiu2023, Silva2019}, which will be the subject of future investigations.

While the peak of prompt fluorescence remains constant due to its temperature independence, the delayed fluorescence peak shifts to smaller rates. This can be explained by the reduced thermal activation of the RISC process. Furthermore, the contributions of faster rates play a more significant role for lower temperatures, as the distribution of the decay rates becomes broader. This could be attributed to the limited rearrangement possibilities of molecules at lower temperatures, where the molecular conformations are "frozen" within their environment. 

\begin{figure}[H]
    \centering
    \includegraphics[width=8 cm]{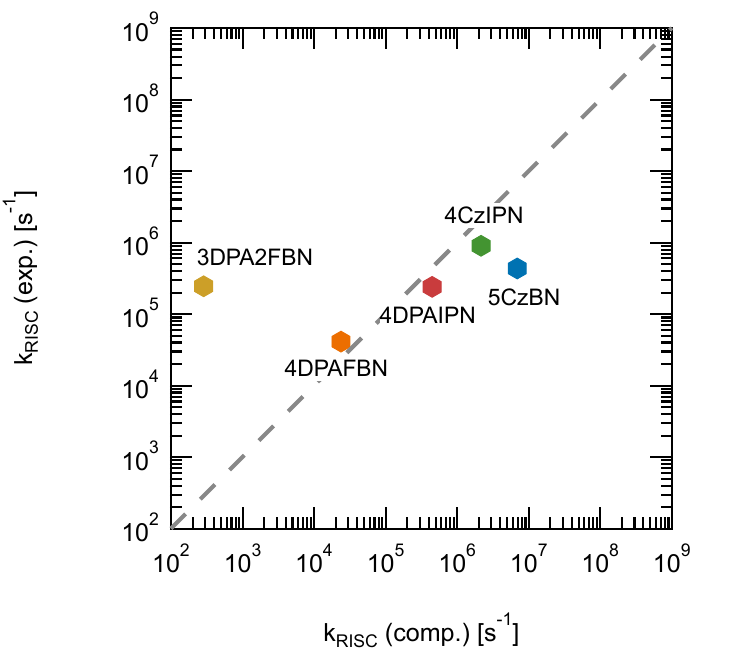}
    \caption{Scatter plot of experimentally and computationally determined reverse intersystem crossing rates for all molecules.}
    \label{fig:rate_scatter}
\end{figure} 

The experimental values for $k_\mathrm{RISC}$ obtained with the 'Gamma-Fit' method match the ones obtained with TD--DFT only with varying degrees of consistency, as shown in Figure \ref{fig:rate_scatter} (and in Tables S1, S3, and S4 in the Supporting Information). To evaluate the predictive reliability of our computational modeling results across the different molecules, we performed a multivariate statistical sensitivity analysis using a linear mixed model with the range-normalized absolute error (RNAE) between the experimentally derived values and the computational values as the dependent variable. \cite{lindstrom1988newton} We employed several predictors in order to gauge the potential origins of this varying deviation, such as the use of a Marcus-like approach, the employed level of theory, and both the diversity of molecular conformations as well as the diversity of (R)ISC pathways, represented by a Shannon index and a Pathway index, respectively. The modeling details can be found in Section S9 of the Supporting Information.

Our evaluation reveals that the primary predictor of computational modeling inaccuracy is the donor substituent class ($p = 0.037$) rather than the specific functional or semiclassical computational framework used. We identified a decrease in computational accuracy between rigid, Cz-based emitters and flexible, DPA-substituted analogues. While our computational approach maintains good accuracy for comparatively rigid emitters like 5CzBN (and others, see above), the DPA-based molecules exhibit a systematic increase in RNAE ($\beta = 0.952$). This discrepancy mirrors our experimental observations where the increased degrees of freedom in DPA units lead to faster non-radiative decay from the excited singlet state and consequently lower PLQYs. Physically, this suggests that a static geometry approximation, meaning calculations on one static, optimized structure, is insufficient for flexible systems where the conformational ensemble deviates significantly from a single minimum. The same conformational space that facilitates non-radiative processes in the experimental setup introduces a structural penalty in the static calculations, as a single optimized minimum geometry cannot represent the dynamic ensemble of a flexible molecule. This is underlined by the fact that improving the level of theory from a DFT hybrid approach to an optimally tuned range-separated hybrid approach does not lead to a consistent improvement of results on the same geometry, despite the more accurate description of the essential CT states. \cite{stein2009reliable} We also observed that the accessibility of RISC pathways mediated by higher triplet states impacts the computational modeling error (see Sections S9.3 and S9.4 of the Supporting Information). This further indicates that the sensitivity of electronic couplings to small geometric fluctuations is additionally amplified in more flexible molecular architectures. The importance of $T_\mathrm{2}$-mediated RISC pathways was also previously pointed out by Aizawa et al. \cite{Aizawa2020}

\begin{figure}[H]
    \centering
    \includegraphics[width=10cm]{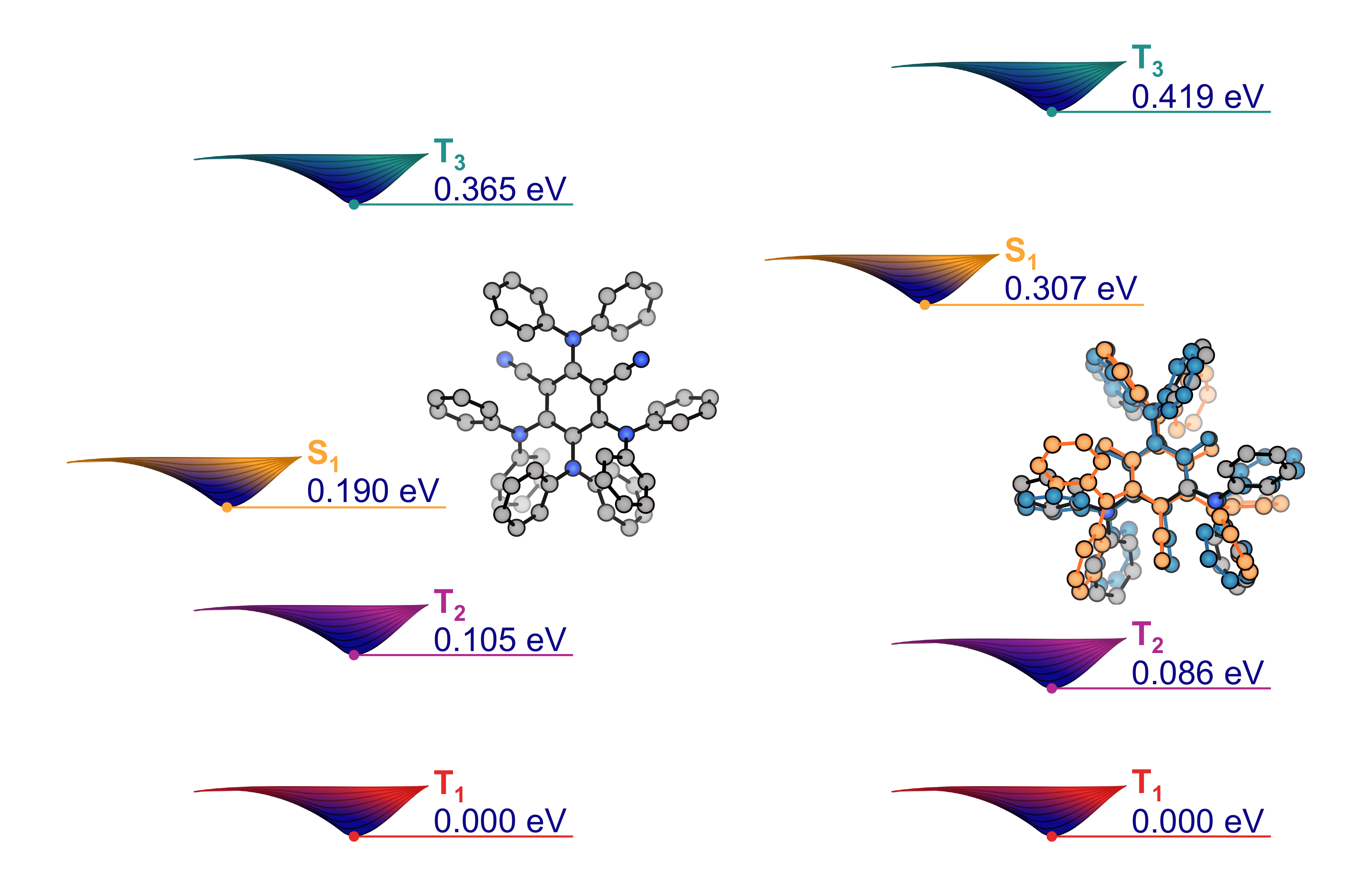}
    \caption{Adiabatically corrected singlet and triplet energies for the global minima of 4DPAIPN (left) and 3DPA2FBN (right), relative to the respective lowest triplet state (for details on the adiabatically corrected energies see Section S9.1 of the Supporting Information). Additional conformers within $3 \ \mathrm{kcal} \ \mathrm{mol^{-1}}$ of the respective global $S_\mathrm{0}$  minima are shown in different colors. 4DPAIPN has no additional conformers within this range, while 3DPA2FBN has two.}
    \label{fig:energy_level}
\end{figure}

However, the error discrepancy between Cz-based and DPA-based emitters is gradual and not solely a function of the substituent type, but also of the steric environment in which it resides: In molecules such as 4DPAIPN and 4DPAFBN, the higher density of DPA groups leads to spatial proximity that sterically "locks" the substituents (see Figure \ref{fig:energy_level}). This limits the accessible conformational space, making the optimized global minimum geometry a more reliable approximation.

The computational accuracy is limited most significantly in 3DPA2FBN: Despite possessing fewer DPA groups, the replacement of a bulky DPA unit with a smaller fluorine atom as compared to 4DPAFBN provides the necessary spatial volume for the remaining substituents to fully leverage their additional degrees of freedom. This increased flexibility results in a broader distribution of accessible conformers (see Figure \ref{fig:energy_level}). Since the computational protocol relies on a static geometry, it fails to capture the fluctuations of photophysical properties inherent to this broader conformational ensemble.

Interestingly, the experimental non-radiative rates do not correlate well with the calculated $S_1$--$S_0$ energy gap expected from the energy gap law. For instance, while 4DPAFBN exhibits an increased energy gap relative to 4DPAIPN ($+ \ 0.13\,\mathrm{eV}$), it displays higher non-radiative rates, suggesting that non-radiative loss is not solely dependent on the number of available quenching centers (DPA units). This can also be observed when comparing 4DPAFBN to 3DPA2FBN, where the magnitude of non-radiative processes decreases. Even though the remaining three DPA groups are allotted more space for movement, the total sum of the non-radiative pathways is lower, simply because there are fewer overall DPA vibrational modes to dissipate the energy compared to a molecule with four DPA substituents.

Overall, our analysis revealed that the predictive reliability of the computational modeling approach is primarily governed by the structural flexibility of the donor substituents rather than the specific electronic structure method or theoretical framework. While static, equilibrium-based calculations perform well for rigid architectures, they struggle to represent the multi-conformational landscape of flexible DPA-substituted systems, where a single optimized minimum fails to capture the dynamic ensemble. However, this discrepancy is highly sensitive to the local steric environment, which can either constrain these degrees of freedom through crowding of substituents or permit the fluctuations that undermine the static modeling. Because these limitations are geometric in origin, even transitioning to more sophisticated functionals cannot rectify the error if the underlying model remains rooted in a single static geometry. 
Additionally, while structural flexibility sets the baseline for computational modeling error, the inclusion of $T_\mathrm{n}$-mediated pathways can still be employed to improve computational accuracy and should not be neglected (see Figure \ref{fig:energy_level}).

\section*{Conclusion}\label{sec13}

In summary, we presented the new 'Gamma-Fit' method as a robust, straightforward and physically intuitive framework for modeling the multiexponential TADF decays that are characteristic of disordered thin films. By accounting for a continuous distribution of decay rates rather than discrete exponential components, this approach captures the inherent structural heterogeneity of solid-state environments.

Our investigation into a series of TADF emitters revealed that substitution of the carbazole donor groups with DPA donors reduces the delayed fluorescence due to high non-radiative losses and a substantial increase in $\Delta E_\mathrm{ST}$ exceeding $100\,\mathrm{meV}$. Therefore, non-flexible donor groups are preferable in TADF applications. Our results emphasize that the local environment remains a dominant factor in determining the overall TADF efficiency. For this purpose, the newly introduced 'Gamma-Fit' method offers an accurate and practical solution. 

Furthermore, our study interrogates the limits of commonly employed computational approaches. We demonstrated that our computational protocol yields reliable results for rigid molecules but reaches a computational accuracy limit when the assumption of structural rigidity inherent to the carbazole-based species is applied to systems containing more flexible substituents, as shown for DPA donors. While single-conformation models provide a useful baseline, a more predictive model must account for the ensemble of molecular geometries and the contribution of higher-lying, RISC-active triplet states. This provides a clear incentive for adopting ensemble-averaging or dynamic sampling for flexible TADF emitters combined with a more thorough exploration of their excited state potential energy surfaces.

\section*{Experimental and Computational Details}

\subsection*{Experimental Setup}

Thin film samples were prepared by drop coating. Glass substrates were used and cleaned with ethanol, acetone and isopropanol in an ultra sonic bath. Pure TADF samples were dissolved in toluene at a mass concentration of $0.1\,\mathrm{mg\,ml^{-1}}$. 

Time resolved photoluminescence decays were measured with a self-built confocal microscope in a Linkam cryostat (LTS420) with a Thorlabs super apochromatic microscope objective ($10\times,~\mathrm{NA}=0.5$). The decays were detected with time-correlated single-photon counting (TCSPC) using a Picoquant PMA Hybrid 40 detector and a PicoHarp300 TCSPC module. A time resolution of $512\,\mathrm{ps}$ was selected on the PicoHarp300 in order to achieve a measurement range of $33\,\mathrm{\mu  s}$. 
The samples were excited at $405\,\mathrm{nm}$ using the femtosecond laser Pharos 20 W from Light Conversion ($IRF < 1\,\mathrm{ps}$) at a repetition rate of $25\,\mathrm{kHz}$. Fluorescence was detected using a $450\,\mathrm{nm}$ long-pass filter. Each decay curve was measured by integrating for $15\,\mathrm{min}$.
The data was background-corrected, logarithmically binned, and normalized to the maximum value. The temperature-dependent data sets were then fitted globally according to eq (\ref{eq_Gamma_TADF}) and the decay parameters were extracted.

The photoluminescence quantum yield at room temperature was approximated using an integrating sphere (Ocean Optics, ISP-80-8-R) and the QEPro Ocean Optics spectrometer according to the method of de Mello et al. \cite{Mello1997} Both devices were coupled using the QP600-2-UV-BX fibre. The molecules were excited at $405\,\mathrm{nm}$.

\subsection*{Quantum-Chemical Calculations}

\subsubsection*{Conformational Sampling and Geometry Optimization}\label{sec:QM1}

\noindent Geometries of all molecules were constructed using GaussView 6. \cite{gv6} Subsequent ground state optimizations were carried out with Gaussian 16, \cite{g16} using the B3LYP functional \cite{vosko1980accurate, lee1988development, becke1988density, becke1993density, stephens1994ab} augmented by Grimme’s D3 dispersion correction \cite{grimme2010consistent} in conjunction with the 6-31+G* Pople basis set. \cite{ditchfield1971a, hehre1972a, hariharan1973a, francl1982a, gordon1982a, clark1983a, spitznagel1987a} This level of theory is based on our previous studies of carbazole-containing cyanoarenes as TADF emitter materials. \cite{Streiter2020, morgenstern2025unlocking} All resulting optimized structures were confirmed as minimum energy geometries by vibrational frequency analysis.\\
To explore the ground state conformational space of the cyanoarene systems, the global optimizer algorithm (GOAT) module \cite{de2025goat} implemented in Orca 6 \cite{ORCA, neese2025software} was used in combination with the semiempirical GFN2-XTB method. \cite{bannwarth2019gfn2} The global minima of each molecule served as starting points for TD--DFT optimizations of the first singlet ($S_\mathrm{1}$) and triplet ($T_\mathrm{1}$) excited states with the same level of theory used for the ground state optimizations.

\subsubsection*{Computational Determination of Photophysical Properties}\label{sec:QM2}

All electronic energies and SOC matrix elements $H_\mathrm{SO}$ were calculated at the corresponding optimized geometries using Orca 6 and the B3LYP-D3/def2-SVPD \cite{weigend2005a, rappoport2010a} level of theory with TD--DFT as well as an optimally tuned (OT) version of the range-separated hybrid functional $\omega$B97X-D3 \cite{lin2013long} with the same specifications to improve upon the description of CT states. The Tamm-Dancoff approximation was employed for triplet calculations with  OT-$\omega$B97X-D3. Range separation parameters $\omega$ were determined for the ground state global minimum of each molecule by minimizing the following function \cite{stein2009reliable} derived from Koopmans' theorem \cite{koopmans1934zuordnung} using Brent’s algorithm: \cite{brent1971algorithm}

\begin{equation}
    J^2(\omega) = \sum\limits_{i=0}^{1}(\varepsilon_\mathrm{HOMO}(N+i)+IP(N+i))^2
\end{equation}
\\
$\varepsilon_\mathrm{HOMO}$ and $IP$ are the HOMO energies and the ionization potentials of the neutral $N$-electron and the anionic $(N+1)$-electron systems. The obtained molecule-specific optimal range separation parameters are given in Table S2.\\
\noindent(R)ISC rates between an initial state $i$ and a final state $f$ were determined using a semiclassical Marcus-like approach: \cite{marcus1993electron, bredas2004charge, Olivier2017}

\begin{equation}
    k_{i\rightarrow f} = \frac{2\pi}{\hbar}|H_\mathrm{SO}|^2 \frac{1}{\sqrt{4\pi \lambda_{i\rightarrow f}k_\mathrm{B}T}}\exp{-\frac{E_\mathrm{A}}{k_\mathrm{B}T}}
\end{equation}
\\
$\hbar$ is the reduced Planck constant, $k_\mathrm{B}$ is the Boltzmann constant, and $T$ is the temperature. The activation energy $E_\mathrm{A}$ is determined as follows:

\begin{equation}
    E_\mathrm{A} = \frac{(\Delta E_{i\rightarrow f}+\lambda_{i\rightarrow f})^2}{4\lambda_{i\rightarrow f}} \ ,
\end{equation}
\\
with the adiabatic singlet--triplet gap $\Delta E_{i\rightarrow f}$:

\begin{equation}
    \Delta E_{i\rightarrow f} = E_f - E_i
\end{equation}
\\
$E_f$ and $E_i$ are the energies of the final and initial states, respectively. The sign of $\Delta E_{i\rightarrow f}$ is usually negative for ISC and positive for RISC due to the relative position of the activation energy barrier in relation to the respective initial state.\\
The reorganization energy $\lambda_{i\rightarrow f}$ was obtained as the difference between the final state energy at the initial geometry and at the final geometry:

\begin{equation}
    \lambda_{i\rightarrow f} = E_f(i) - E_f(f)
\end{equation}
\\
This represents the energy required for the molecular geometry to relax into the final state following the electronic transition, since the electronic crossing timescale precludes simultaneous geometric adaptation.\\
Consequently,  $H_\mathrm{SO}$ is computed at the optimized $S_\mathrm{1}$ geometry for ISC and at the optimized $T_\mathrm{1}$ geometry for RISC. For ISC, $H_\mathrm{SO}$ is calculated as the sum of the contributions of all three triplet sublevels:

\begin{equation}
    |H^\mathrm{S}_\mathrm{SO}|^2 = \sqrt{|H^\mathrm{x}_\mathrm{SO}|^2+|H^\mathrm{y}_\mathrm{SO}|^2+|H^\mathrm{z}_\mathrm{SO}|^2}
\end{equation}
\\
For RISC, it is calculated as the average, since each triplet sublevel has a Boltzmann weight of $\frac{1}{3}$:

\begin{equation}
    |H^\mathrm{T}_\mathrm{SO}|^2 = \sqrt{\frac{|H^\mathrm{x}_\mathrm{SO}|^2+|H^\mathrm{y}_\mathrm{SO}|^2+|H^\mathrm{z}_\mathrm{SO}|^2}{3}}
\end{equation}
\\
All properties calculated with this approach can be found in Tables S3 and S4 (see Supporting Information).\\

\noindent A detailed breakdown of the statistical modeling can be found in Section S9 of the Supporting Information. Evaluation of triplet CT character for the construction of the Pathway index predictor in Section S9.1 was carried out with TheoDORE 3.2, \cite{plasser2020theodore} partitioning each molecule into a single donor fragment containing all Cz or DPA moieties, and an acceptor fragment containing the remaining part of the molecule. Molecular structures were rendered with xyzrender. \cite{Goodfellow2026}

\section*{Acknowledgements}

This work was funded by the German Research Foundation (DFG), TRR-386, TPs B06 and B08, project number 514664767. The authors gratefully acknowledge the resources on the LiCCA HPC cluster of the University of Augsburg, co-funded by the DFG – Project-ID 499211671.

\section*{Supporting Information}

The following files are available free of charge.
\begin{itemize}
  \item Supporting Information: General Methods, General Procedures for TADF Emitter Synthesis, $^1$H NMR Spectra of TADF Emitters, Additional Fitting Details, Calculations of the TADF Parameters, Laplace Transform of the Gamma Distribution, Excitation and Photoluminescence Spectra of thin film samples, Computational Data, and Statistical Modeling
\end{itemize}

\printbibliography
\end{refsection}
\clearpage

\appendix
\begin{refsection}
{
    \centering
    {\LARGE \textbf{Supporting Information}\\Beyond the Static Approximation: Assessing the Impact of Conformational and Kinetic Broadening on the Description of TADF Emitters\\}
    \vspace{0.5cm}
    {\large Daniel Beer$^1$, Jonas Weiser$^2$, Tom Gabler$^3$, Kirsten Zeitler$^3$, Carsten Deibel$^1$, Christian Wiebeler$^{*2}$\\}
    \vspace{0.5cm}
    {\large $^1$ Institut für Physik, Technische Universität Chemnitz, 09126 Chemnitz, Germany\\ $^2$ Institut für Physik \& Center for Advanced Analytics and Predictive Sciences, Universität Augsburg, 86159 Augsburg, Germany\\ $^3$ Institut für Organische Chemie, Universität Leipzig, 04103 Leipzig, Germany\\}
    \vspace{0.5cm}
    \date{\large *Email: christian.wiebeler@uni-a.de\\}
}

\tableofcontents
\clearpage

\setcounter{secnumdepth}{1}
\setcounter{secnumdepth}{2}

\renewcommand{\thesection}{S\arabic{section}}
\renewcommand{\thesubsection}{S\arabic{section}.\arabic{subsection}}
\renewcommand{\thetable}{S\arabic{table}}
\renewcommand{\thefigure}{S\arabic{figure}}

\section{General Methods}

\subsection{Solvents and Reagents}

The syntheses were carried out using standard laboratory borosilicate glass equipment. All reactions and working steps were conducted under argon in the absence of air and water using standard Schlenk techniques. Solvents used for synthesis were of p.a. quality. Solvents for chromatography (Dichloromethane (DCM), Petroleum ether (PE)) were technically pure and distilled before use. Tetrahydrofuran (THF) was distilled over elemental sodium and benzophenone. Purchased substances were used without further purification unless otherwise stated.

\subsection{Thin layer chromatography (TLC)}

Thin layer chromatographic studies were carried out using TLC prefabricated films \textsc{Alugram}$^\text{\faRegistered[regular]}$ \textit{Xtra SIL G/UV$245$} from \textsc{Macherey-Nagel} of the silica gel 60 F UV245 type with a layer thickness of $0.20\,\mathrm{mm}$ silica gel. UV light ($254\,\mathrm{nm}$ and $366\,\mathrm{nm}$) was used for detection.

\subsection{Column chromatography (CC)}

Colum chromatographic separations were performed using \textsc{Biotage} \textit{Isolera One} automatic chromatography instrument with flash silica gel type Silica 60 (particle size: $25-40\,\mathrm{\mu m}$) from \textsc{Macherey-Nagel}. Fractions were detected using a UV detector at $254\,\mathrm{nm}$ and $280\,\mathrm{nm}$.

\subsection{NMR spectroscopy}

NMR spectra were recorded on either a Bruker \textit{AVANCE III HD $400$} ($^1$H: 300 MHz), a \textsc{Varian} \textit{MERCURYplus $400$} ($^1$H: $400\,\mathrm{MHz}$) or a \textsc{Varian} \textit{MERCURYplus $300$} ($^1$H: $300\,\mathrm{MHz}$) spectrometer. All spectra were recorded at room temperature. Chemical shifts are in ppm ($\delta$-scale) and coupling constants are in Hertz (Hz). In $^1$H spectra, the chemical shift $\delta$ was referenced to the respective residual proton signal of the solvent (CHCl$_3$: $\delta(^1\text{H}) = 7.26\,\mathrm{ppm}$, DMSO: $\delta(^1\text{H}) = 2.50\,\mathrm{ppm}$, DCM: $\delta(^1\text{H}) = 5.32\,\mathrm{ppm}$). \cite{Fulmer2010} The spectra obtained were processed and analyzed using \textit{MestReNova} software (version 14.1.0) from \textsc{Mestrelab Research} S.L. The following abbreviations are used to indicate the multiplicity of signals: s (singlet), d (doublet), t (triplet), m (multiplet). For higher order, multiplet combinations of already described abbreviations were used: dd (doublet of doublets), ddd (doublet of doublets of doublets), dt (doublet of triplets). The frequencies and solvents used are indicated in each case with the spectroscopic data in the experimental procedures.

\section{General Procedure for TADF Emitter Synthesis}\label{sec:TADF_synthesis}

TADF emitters were synthesized according to previous work. \cite{Speckmeier2018, Garreau2019, Chernowsky2021} \\ \\ 
In a flame dried Schlenk flask, car\-bazole or di\-phenyl\-amine ($1.25$ equiv per halogen of the benzene car\-bo\-nitrile or iso\-phthalo\-nitrile) was dissolved in dry THF ($0.05\,$\textsc{M}, relating to $1.00$ equiv of the corresponding cyano\-benzene) under an argon atmosphere. Then, $1.88$ equiv (per halogen of the cyano\-benzene) NaH ($60\,\%$ in mineral oil) were added and the suspension was stirred at room temperature for $1\,\mathrm{h}$. With di\-phenyl\-amine, the suspension was instead stirred at $50\,^\circ\mathrm{C}$ for $30$ minutes. Finally, $1.00$ equiv of the fluorinated cyanoarene or iso\-phthalo\-nitrile was added and the resulting mixture was stirred at room temperature for $15$ to $24\,\mathrm{h}$. The reaction mixture was quenched by the addition of water ($2\,\mathrm{mL}$). After removal of THF, the residue was dissolved in DCM and washed with water. The organic phase was dried over $\mathrm{Na_2SO_4}$ and the solvent was removed under reduced pressure. The crude product was purified by flash chromatography on silica gel.
\\ \\
\clearpage

\noindent\textbf{5CzBN; 2,3,4,5,6-Penta(9H-carbazol-9-yl)benzonitril}\\ \\
\begin{minipage}{0.4\textwidth}
    \centering
    \includegraphics[width=0.5\linewidth]{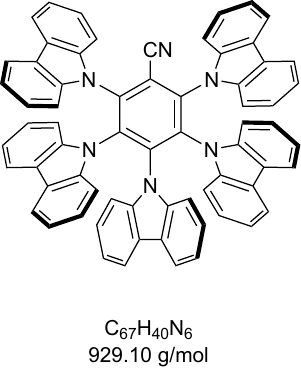}
\end{minipage}
\begin{minipage}{0.6\textwidth}
    According to the general procedure using $194\,\mathrm{mg}$ penta\-fluoro\-benzo\-nitrile ($1.0\,\mathrm{mmol}$, $1.0$ equiv) and $20\,\mathrm{ml}$ THF.Yield after flash chromatography (PE/DCM, $15-80\,\%$ DCM): $799\,\mathrm{mg}$ ($0.83\,\mathrm{mmol}$, $83\,\%$), yellow solid.\\ \\
    \textbf{$^1$H}$\:$\textbf{NMR} ($400\,\mathrm{MHz}$, DMSO-d$_6$): $\delta\, = 7.91\,$--$\,7.80$ (m, 8H), $7.78\,$--$\,7.68$ (m, 6H), $7.41\,$--$\,7.35$ (m, 4H), $7.36\,$--$\,7.29$ (m, 2H), $7.18\,$--$\,7.11$ (m, 4H), $7.11\,$--$\,7.04$ (m, 4H), $6.78\,$--$\,6.71$ (m, 4H), $6.71\,$--$\,6.63$ (m, 6H), $6.62\,$--$\,6.55$ (m, 2H).\\ \\
    All data in accordance to the literature. \cite{Tanimoto_2016}

\end{minipage}
\vspace{1.5em}

\textbf{4CzIPN; 2,4,5,6-Tetrakis(9\textit{H}-carbazol-9-yl)isophthalonitrile}\\ \\ 
\begin{minipage}{0.4\textwidth}
    \centering
    \includegraphics[width=0.5\linewidth]{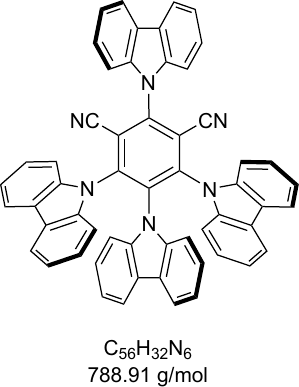}
\end{minipage}
\begin{minipage}{0.6\textwidth}
    According to the general procedure using $204\,\mathrm{mg}$ tetra\-fluoro\-iso\-phthalo\-nitril ($1.0\,\mathrm{mmol}$, $1.0$ equiv) and $20\,\mathrm{ml}$ THF. Yield after flash chromatography (PE/DCM, $15-60\,\%$ DCM): $582\,\mathrm{mg}$ ($0.74\,\mathrm{mmol}$, $74\,\%$), yellow solid. \\ \\
    \textbf{$^1$H}$\:$\textbf{NMR} ($400\,\mathrm{MHz}$, DMSO-d$_6$): $\delta\, = 8.39\,$--$\,8.32$ (m, 2H), $8.23\,$--$\,8.17$ (m, 2H), $7.89\,$--$\,7.82$ (m, 4H), $7.79\,$--$\,7.71$ (m, 6H), $7.58\,$--$\,7.52$ (m, 2H), $7.52\,$--$\,7.42$ (m, 4H), $7.19\,$--$\,7.06$ (m, 8H), $6.85\,$--$\,6.77$ (m, 2H), $6.75\,$--$\,6.66$ (m, 2H).\\ \\
    All data in accordance to the literature. \cite{Uoyama2012}
\end{minipage}
\vspace{1.5em}

\textbf{4DPAIPN; 2,4,5,6-Tetrakis(diphenylamino)isophthalonitrile} \\ \\
\begin{minipage}{0.4\textwidth}
    \centering
    \includegraphics[width=0.5\linewidth]{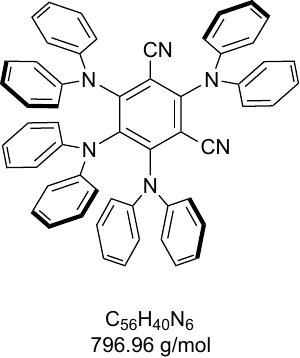}
\end{minipage}
\begin{minipage}{0.6\textwidth}
    According to the general procedure using $204\,\mathrm{mg}$ tetra\-fluoro\-iso\-phthalo\-nitril  ($1.0\,\mathrm{mmol}$, $1.0$ equiv) and $20\,\mathrm{ml}$ THF. Yield after flash chromatography (PE/DCM, $12-90\,\%$ DCM): $654\,\mathrm{mg}$ ($0.82\,\mathrm{mmol}$, $82\,\%$), yellow solid. \\ \\
    \textbf{$^1$H}$\:$\textbf{NMR} ($400\,\mathrm{MHz}$, CDCl$_3$): $\delta\, = 7.30\,$--$\,7.24$ (m, 4H),  $7.11\,$--$\,7.00$ (m, 14H), $6.94\,$--$\,6.84$ (m, 8H), $6.75\,$--$\,6.64$ (m, 10H), $6.59\,$--$\,6.51$ (m, 4H). \\ \\
    All data in accordance to the literature. \cite{Garreau2019, Chernowsky2021} 
\end{minipage} 
\vspace{1.5em}

\textbf{4DPAFBN; 2,3,4,6-Tetra(diphenylamino)-5-fluorobenzonitrile} \\ \\
\begin{minipage}{0.4\textwidth}
    \centering
    \includegraphics[width=0.5\linewidth]{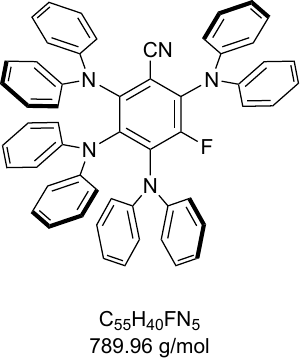}
\end{minipage}
\begin{minipage}{0.6\textwidth}
    According to the general procedure using $386\,\mathrm{mg}$ penta\-fluoro\-benzo\-nitrile  ($2.0\,\mathrm{mmol}$, $1.0$ equiv) and $40\,\mathrm{ml}$ THF. Yield after flash chromatography (PE/DCM, $15-90\,\%$ DCM): $1405\,\mathrm{mg}$ ($1.78\,\mathrm{mmol}$, $89\,\%$), yellow solid. \\ \\
    \textbf{$^1$H}$\:$\textbf{NMR} ($400\,\mathrm{MHz}$, DCM-d$_2$): $\delta\, = 7.32\,$--$\,7.26$ (m, 16H), $7.10\,$--$\,7.04$ (m, 8H), $7.03\,$--$\,6.98$ (m, 16H). \\ \\
    All data in accordance to the literature. \cite{Speckmeier2018}
\end{minipage}
\vspace{1.5em}

\textbf{3DPA2FBN; 2,4,6-Tris(diphenylamino)-3,5-difluorobenzonitrile} \\ \\
\begin{minipage}{0.4\textwidth}
    \centering
    \includegraphics[width=0.5\linewidth]{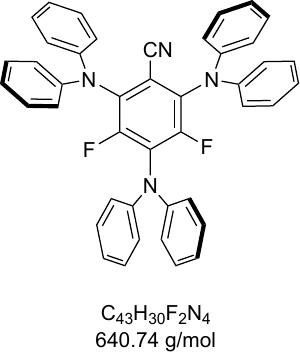}
\end{minipage}
\begin{minipage} {0.6\textwidth}
    According to the general procedure using $193\,\mathrm{mg}$ penta\-fluoro\-benzo\-nitrile  ($1.0\,\mathrm{mmol}$, $1.0$ equiv) and $20\,\mathrm{ml}$ THF. Yield after flash chromatography (PE/DCM, $7-60\,\%$ DCM): $670\,\mathrm{mg}$ ($0.77\,\mathrm{mmol}$, $77\,\%$), yellow solid. \\ \\
    \textbf{$^1$H}$\:$\textbf{NMR} ($400\,\mathrm{MHz}$, DCM-d$_2$): $\delta\, = 7.33\,$--$\,7.22$ (m, 12H), $7.10\,$--$\,7.03$ (m, 6H), $7.03\,$--$\,6.95$ (m, 12H). \\ \\
    All data in accordance to the literature. \cite{Speckmeier2018}
\end{minipage}
\vspace{3em}

\section{$^1$H NMR Spectra of TADF Emitters}\label{sec:NMR_spectra}

\begin{figure}[H]
    \centering
    \includegraphics[width=0.95\linewidth]{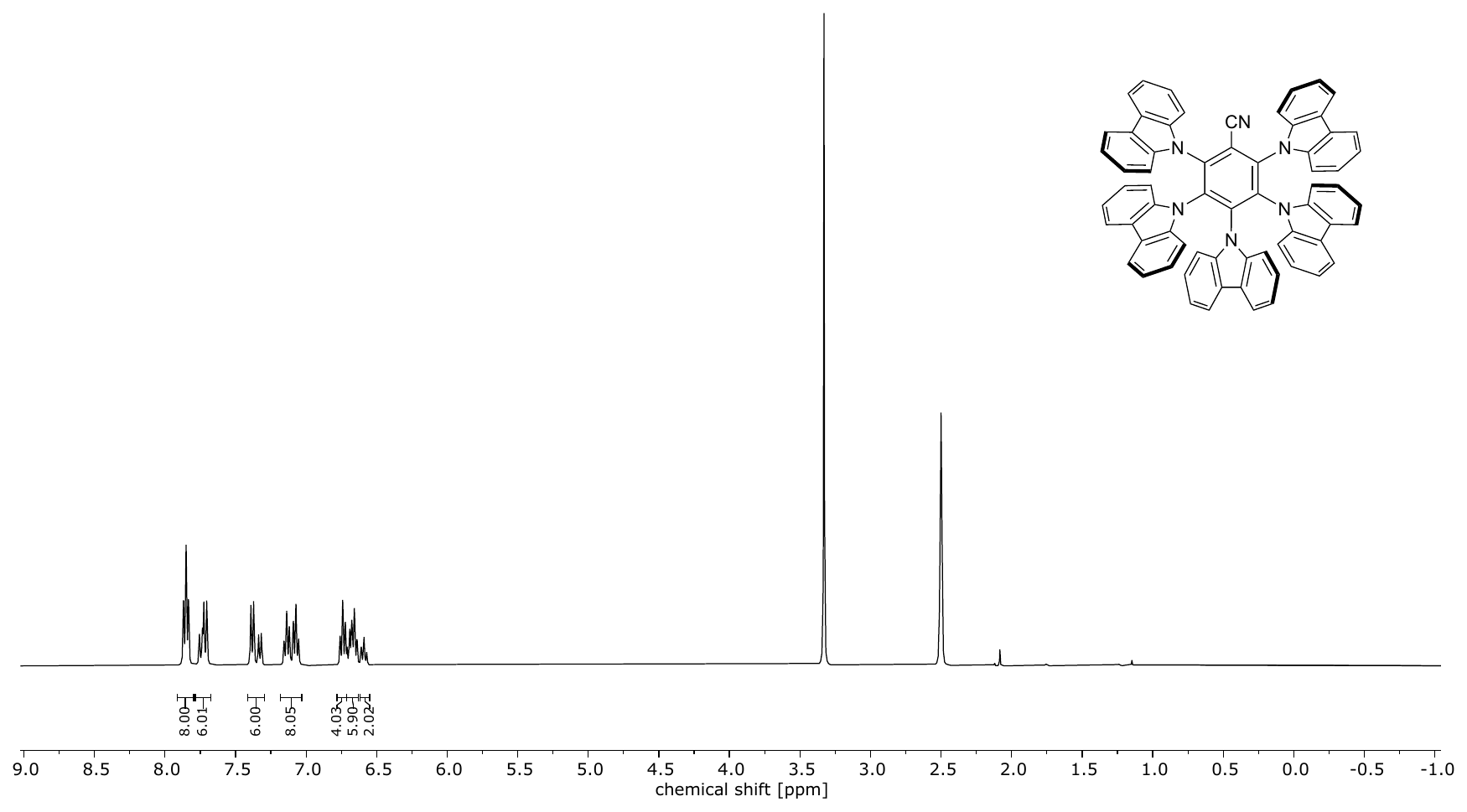}
    \caption{$^1$H NMR spectrum of 5CzBN in DMSO-d$_6$.}
    \label{fig:NMR-5CzBN}
\end{figure} 

\begin{figure}[H]
    \centering
    \includegraphics[width=0.95\linewidth]{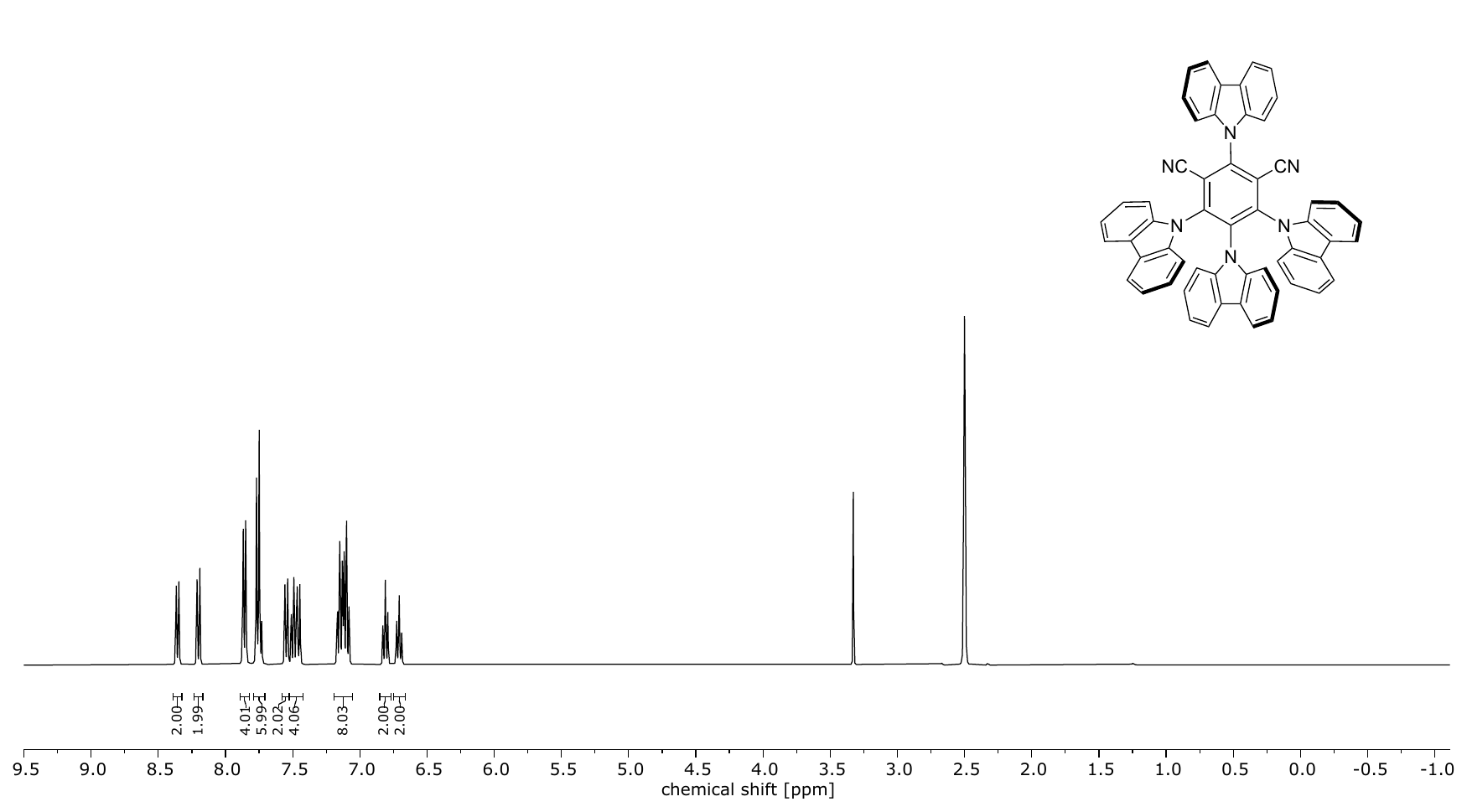}
    \caption{$^1$H NMR spectrum of 4CzIPN in DMSO-d$_6$.}
    \label{fig:NMR-4CzIPN}
\end{figure}

\begin{figure}[H]
    \centering
    \includegraphics[width=0.95\linewidth]{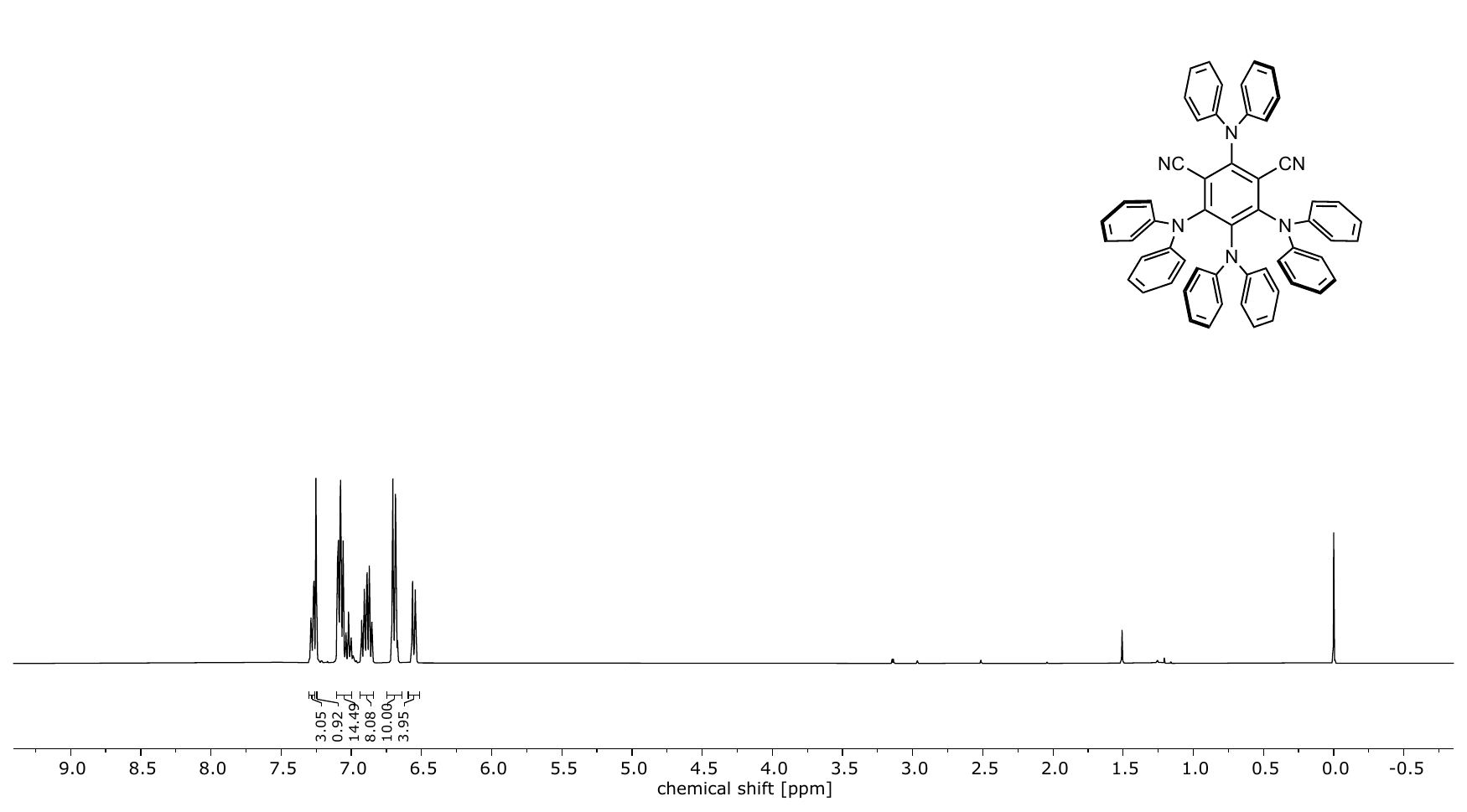}
    \caption{$^1$H NMR spectrum of 4DPAIPN in CDCl$_3$.}
    \label{fig:NMR-4DPAIPN}
\end{figure}

\begin{figure}[H]
    \centering
    \includegraphics[width=0.95\linewidth]{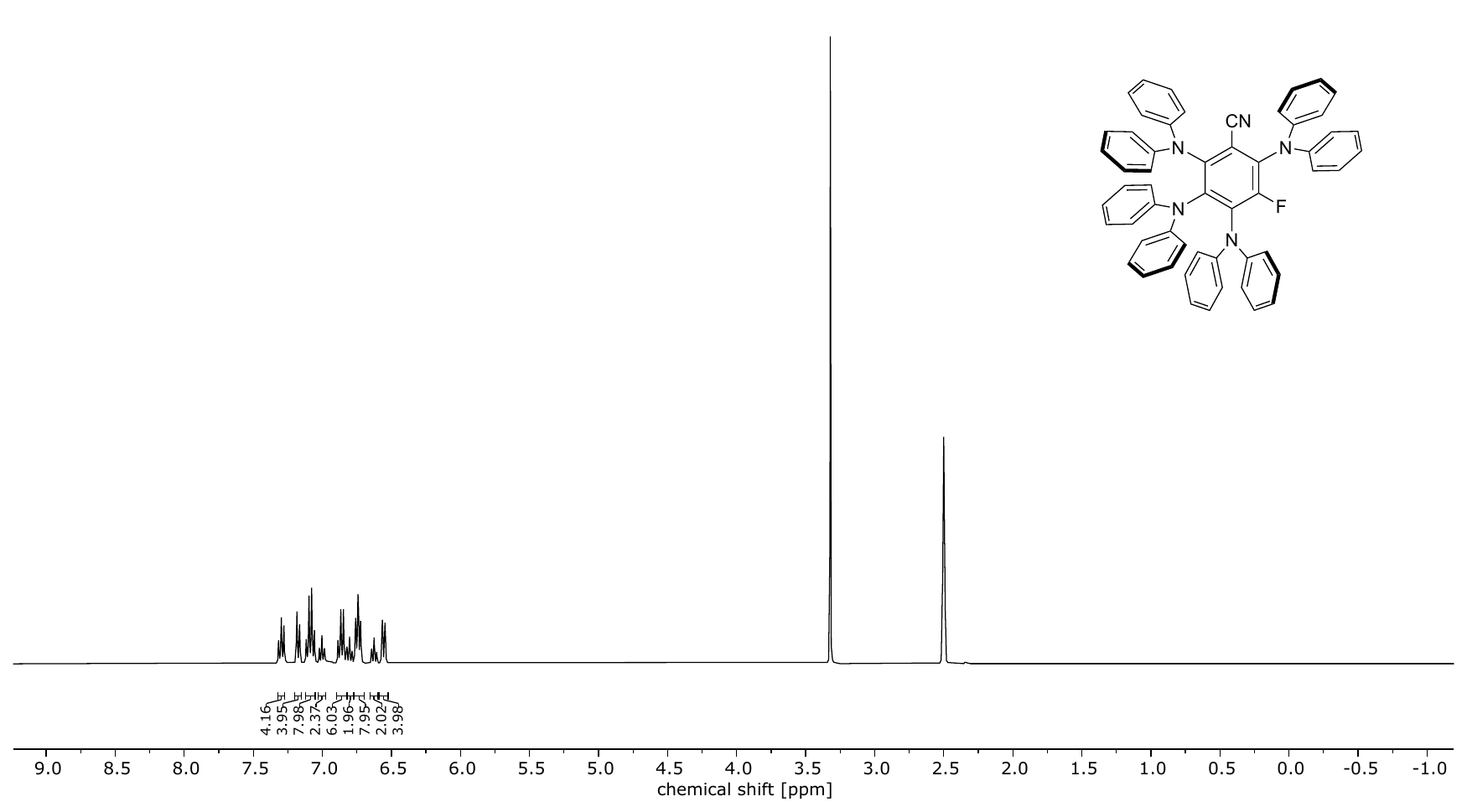}
    \caption{$^1$H NMR spectrum of 4DPAFBN in DCM-d$_2$.}
    \label{fig:NMR-4DPAFBN}
\end{figure}

\begin{figure}[H]
    \centering
    \includegraphics[width=0.95\linewidth]{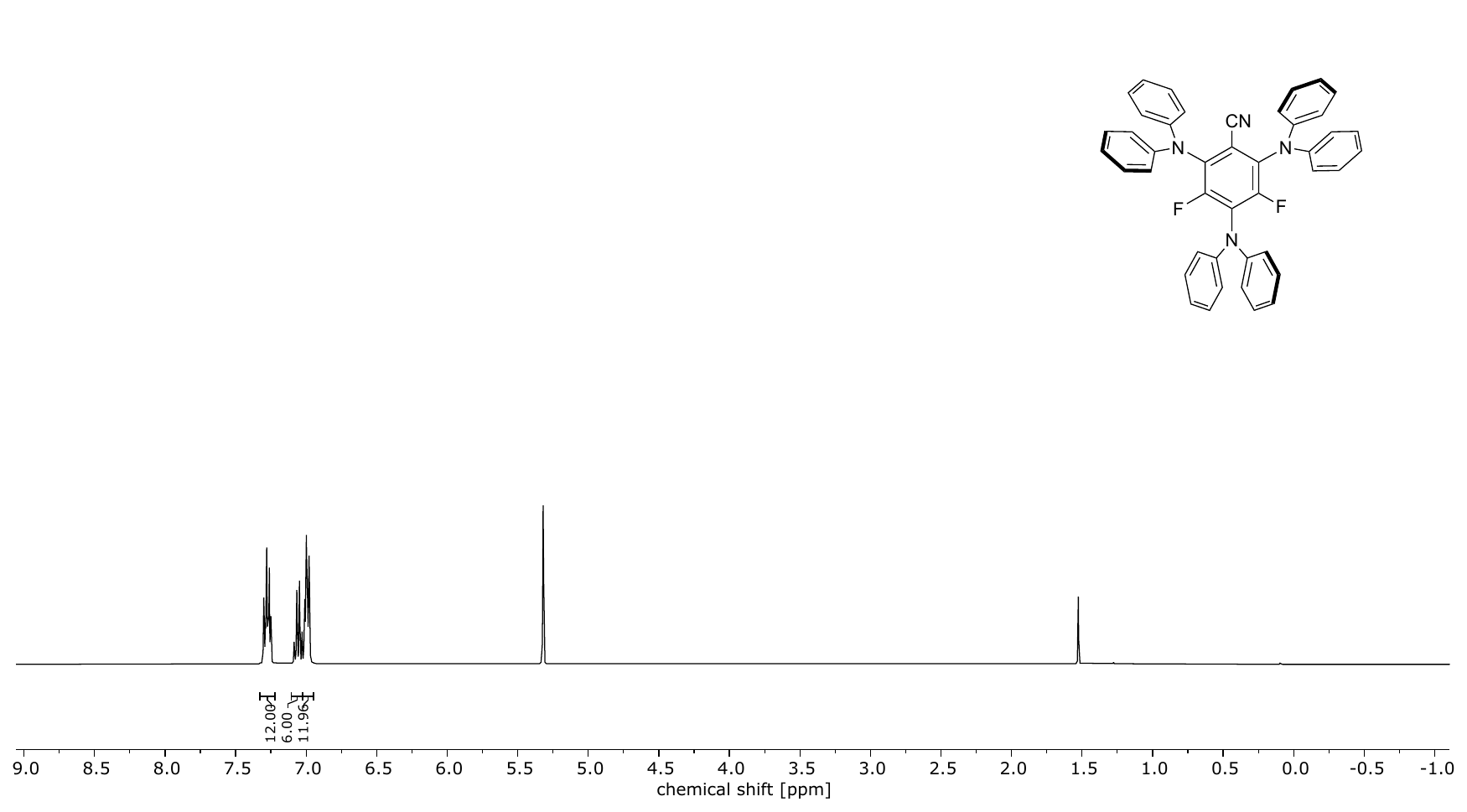}
    \caption{$^1$H NMR spectrum of 3DPA2FBN in DCM-d$_2$.}
    \label{fig:NMR-3DPA2FBN}
\end{figure}
\clearpage

\section{Additional Fitting Details}\label{sec:fitting}

To fit the decay curves, we have introduced the 'Gamma-Fit' method, which consists of one temperature-independent gamma distribution for the prompt fluorescence and one temperature-dependent gamma distribution for the delayed fluorescence. The fit function can be written as follows:

\begin{equation}
    \begin{split}
        f(t, T) =&~ \frac{k_\mathrm{PF}^{r_\mathrm{PF}}}{\Gamma(r_\mathrm{PF})} \cdot t^{r_\mathrm{PF}-1} \cdot \exp{(-k_\mathrm{PF}\cdot t)} \\
        &+ A_\mathrm{DF}(T) \cdot \frac{k_\mathrm{DF}(T)^{r_\mathrm{DF}(T)}}{\Gamma(r_\mathrm{DF}(T))} \cdot t^{r_\mathrm{DF}(T)-1} 
        \cdot \exp{(-k_\mathrm{DF}(T)\cdot t)} \ ,
    \label{eq_Gamma_TADF_supps}
    \end{split}
\end{equation}
\\
where $k_\mathrm{PF}$ is the rate of prompt fluorescence, $r_\mathrm{PF}$ is the shape factor for prompt fluorescence, $k_\mathrm{DF}(T)$ is the temperature-dependent rate of delayed fluorescence, $r_\mathrm{DF}(T)$ is the temperature-dependent shape factor for delayed fluorescence, and $A_\mathrm{DF}(T)$ is the pre-factor for the delayed fluorescence. The temperature-dependent parameters can then be expressed as follows:
\begin{align}
    k_\mathrm{DF}(T) &= k_\mathrm{DF,0} \cdot \exp\left(-\frac{k_\mathrm{DF,temp}}{k_\mathrm{B} T}\right) \\
    r_\mathrm{DF}(T) &= r_\mathrm{DF,0} + \frac{r_\mathrm{DF,temp}}{k_\mathrm{B} T} \\
    A_\mathrm{DF}(T) &= A_\mathrm{DF,0} \cdot \exp\left(-\frac{A_\mathrm{DF,temp}}{k_\mathrm{B} T}\right) \ ,
    \label{eq_temp_fitting_parameter_supps}
\end{align}
with $k_\mathrm{DF,0}$, $r_\mathrm{DF,0}$, $A_\mathrm{DF,0}$ as the corresponding temperature-independent parameters. All TADF decays shown in Figure 2 were fitted globally after eq (\ref{eq_Gamma_TADF_supps}) -- eq (\ref{eq_temp_fitting_parameter_supps}). 
\noindent These parameters were used to calculate the TADF parameters according to the adjusted approach by Tsuchiya \textit{et al.} described in Section \ref{sec:TADF_params}. \cite{Tsuchiya2021}

\section{Calculation of the TADF Parameters}\label{sec:TADF_params}

\begin{figure}[H]
    \centering
    \includegraphics[width=0.6\linewidth]{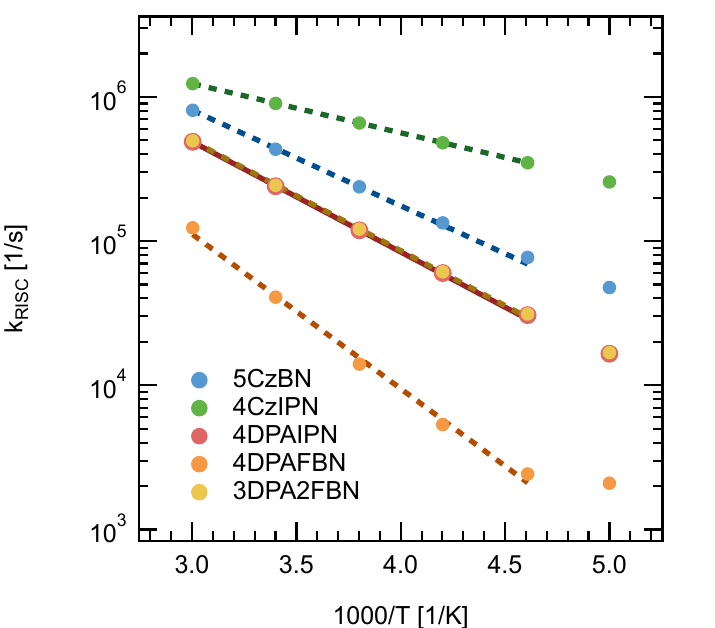}
    \caption{Arrhenius plot of $k_\mathrm{RISC}$ to determine the activation energy $E_\mathrm{A}$ and the decay rate $k_\mathrm{A}$ for all molecules. The data of 4DPAIPN and 3DPA2FBN overlap significantly due to their similar fit parameters.}
    \label{fig:k_risc_temp}
\end{figure}

After global fitting using the temperature dependent curves, the TADF parameters can be calculated according to Tsuchiya \textit{et al.} \cite{Tsuchiya2021} As they assumed an exponential distribution of decay times, small adjustments are necessary to use the 'Gamma-Fit' method. The decay constants for prompt and delayed fluorescence, $k_\mathrm{p}$ and $k_\mathrm{d}$, were determined from the mean values of the gamma distributions: 
\begin{align}
    k_\mathrm{p} &= \mathrm{E}(\Gamma_\mathrm{PF}) = \frac{k_\mathrm{PF}}{r_\mathrm{PF}} \\
    k_\mathrm{d}(T) &= \mathrm{E}(\Gamma_\mathrm{DF}(T) = \frac{k_\mathrm{DF}(T)}{r_\mathrm{DF}(T)}
\end{align}
The efficiencies of prompt ($\Phi_\mathrm{PF}$) and delayed fluorescence ($\Phi_\mathrm{DF}$) were determined from integration of the two gamma distributions over the whole time period:
\begin{align}
    \Phi_\mathrm{PF} &= \int \Gamma_\mathrm{PF}~ dt \\
    \Phi_\mathrm{DF} &= \int \Gamma_\mathrm{DF}~ dt
\end{align}
With this and the neglect of phosphorescence, the equations of Tsuchiya \textit{et al.} can be used to calculate the TADF parameters as follows: \cite{Tsuchiya2021}
\begin{align}
    k_\mathrm{r}^\mathrm{S} =& k_\mathrm{p} \cdot \Phi_\mathrm{PF} \\
    (k_\mathrm{nr}^\mathrm{S})^\mathrm{max} =& k_\mathrm{p} \frac{\Phi_\mathrm{PF}}{\Phi_\mathrm{PLQY}}(1-\Phi_\mathrm{PLQY}) \\
    \begin{split}
            k_\mathrm{ISC}^\mathrm{avg} =& \frac{[k_\mathrm{p}(1-\Phi_\mathrm{PF})-k_\mathrm{d} \Phi_\mathrm{DF}]\Phi_\mathrm{PLQY} + k_\mathrm{p}\Phi_\mathrm{PF}\Phi_\mathrm{DF}}{2\Phi_\mathrm{PF} \Phi_\mathrm{PLQY}}\\ &~\pm \frac{k_\mathrm{p}\Phi_\mathrm{PF}^2(1-\Phi_\mathrm{PLQY})+k_\mathrm{d}\Phi_\mathrm{DF}\Phi_\mathrm{PLQY}}{2\Phi_\mathrm{PF} \Phi_\mathrm{PLQY}}
    \end{split}\\
    (k_\mathrm{nr}^\mathrm{T})^\mathrm{max} =& k_\mathrm{d} \left(1-\frac{\Phi_\mathrm{DF}}{1-\Phi_\mathrm{PF}}\right) \\
    k_\mathrm{RISC}^\mathrm{avg} =& \frac{k_\mathrm{d}}{2} \cdot \frac{\Phi_\mathrm{PLQY}(1-\Phi_\mathrm{PF}) + \Phi_\mathrm{DF} \pm \Phi_\mathrm{PF}(1-\Phi_\mathrm{PLQY})}{\Phi_\mathrm{PF}(1-\Phi_\mathrm{PF})}
\end{align}
Here, $k_\mathrm{r}^\mathrm{S}$ and $k_\mathrm{nr}^\mathrm{S}$ are the respective radiative and non-radiative recombination from the singlet, $k_\mathrm{ISC}$ is the intersystem crossing rate, $k_\mathrm{nr}^\mathrm{T}$ is the rate of the non-radiative recombination from the triplet, and $k_\mathrm{RISC}$ is the reverse intersystem crossing rate. To determine the energy gap between singlet and triplet $\Delta E_\mathrm{ST}$, $k_\mathrm{RISC}$ was calculated for all temperatures and plotted in an Arhenius plot over the inverse temperature.  This is shown in Figure \ref{fig:k_risc_temp}. $E_\mathrm{A}$ and $k_\mathrm{A}$ can be determined from a linear fit of the logarithmic $k_\mathrm{RISC}$ over the inverse temperature according to eq (5).

\begin{figure}[H]
    \centering
    \includegraphics[width=0.6\linewidth]{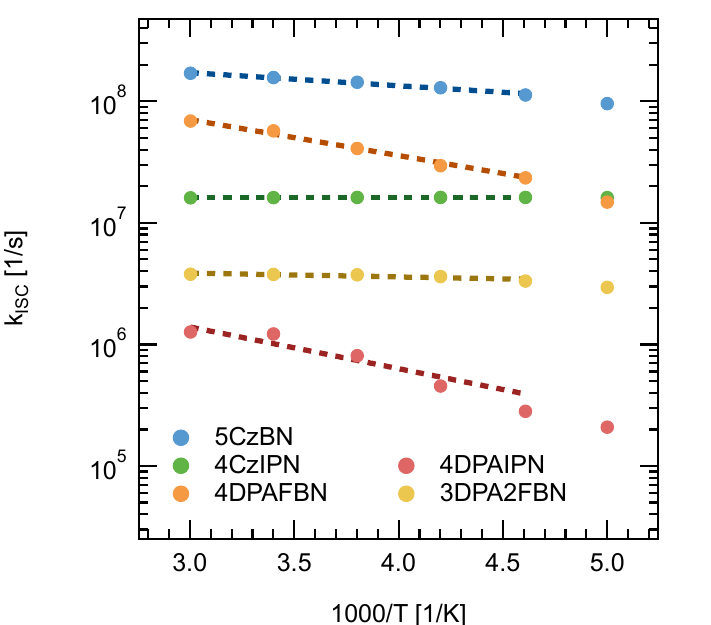}
    \caption{Arrhenius plot of $k_\mathrm{ISC}$ to determine the temperature dependence of the intersystem crossing for all molecules.}
    \label{fig:k_isc_temp}
\end{figure}

\noindent $E_\mathrm{ST}$ can then be calculated from the reorganization energy $\lambda_\mathrm{reorg}$ and $E_\mathrm{A}$ using eq (7). $k_\mathrm{ISC}$ was also identified as temperature dependent. However, this will not be discussed further in this paper. The corresponding dependencies are shown in Figure \ref{fig:k_isc_temp}.

\begin{table}[H]
    \centering
    \caption{Extracted TADF parameters for all emitters at room temperature determined as described in previous sections.}
    \renewcommand{\arraystretch}{1.8}
    \vspace{1em}
    \begin{tabular}{m{3.5em}||m{5.5em}|m{5.5em}|m{5.5em}|m{5.5em}|m{5.5em}}
        \hline
        &5CzBN  & 4CzIPN &4DPAIPN& 4DPAFBN & 3DPA2FBN \\[2ex]
        \hline
        \hline
        $k_\mathrm{r}^\mathrm{S}$ $[10^6\,\mathrm{s^{-1}}]$&$108$ &  $28.0$&$65.9$& $63.3$& $61.4$\\[2ex]
        \hline
        $k_\mathrm{nr}^\mathrm{S}$ $[10^6\,\mathrm{s^{-1}}]$&$37.1$ &  $8.36$&$88.9$& $438$& $90.8$\\[2ex]
        \hline
        $k_\mathrm{nr}^\mathrm{T}$ $[10^4\,\mathrm{s^{-1}}]$&$8.28$ &  $30.5$&$23.6$& $3.61$&$23.4$\\[2ex]
        \hline
        $k_\mathrm{ISC}$ $[10^6\,\mathrm{s^{-1}}]$&$157$ &  $16.1$&$1.21$& $57.1$& $3.76$\\[2ex]
        \hline
        $k_\mathrm{RISC}$ $[10^4\,\mathrm{s^{-1}}]$&$43.3$ &  $90.1$&$24.0$& $4.08$& $24.4$\\[2ex]
        \hline
        $k_\mathrm{A}$ $[10^6\,\mathrm{s^{-1}}]$&$64.3$ & $13.1$&$77.3$& $185$& $86.1$\\[2ex]
        \hline
        $E_\mathrm{A}$ $[\mathrm{meV}]$&$126$ & $67.7$&$146$& $213$& $148$\\[2ex]
        \hline
        $\lambda_\mathrm{reorg}$ [$\mathrm{meV}$] &$188$  & $130$ &$131$& $141$ & $30.2$\\[2ex]
        \hline
        $E_\mathrm{ST}$ $[\mathrm{meV}]$&$120$ & $57.7$&$146$& $205$& $104$\\[2ex]
        \hline        
        $H_\mathrm{SO}$ $[\mathrm{cm^{-1}}]$&$0.32$ & $0.13$&$0.33$& $0.51$& $0.24$\\[2ex]
        \hline
        $\Phi_\mathrm{PF}~$  $[\%]$ &$35.8$ & $53.1$&$42.3$& $11.3$& $39.4$\\[2ex]
        \hline
        $\Phi_\mathrm{DF}$ $[\%]$ &$38.7$ & $23.9$&$0.33$& $1.29$& $0.97$\\[2ex]
        \hline
        $PLQY$ $[\%]$ &$74.5$  & $77.0$ &$42.6$& $12.6$& $40.4$\\[2ex]
        \hline
    \end{tabular}
    \label{tab:4Cz_5Cz}
\end{table}

\begin{figure}[H]
    \centering
    \includegraphics[width=0.5\linewidth]{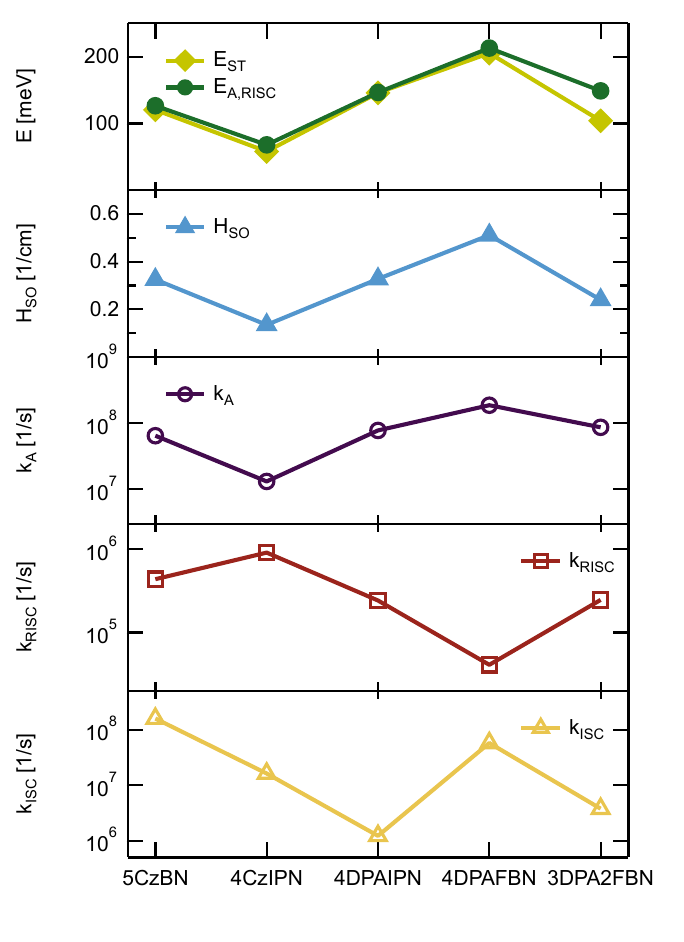}
    \caption{Activation energies $E_\mathrm{A,RISC}$ and singlet-triplet gaps $E_\mathrm{ST}$ of all five molecules compared to the determined spin-orbit couplings $H_\mathrm{SO}$ as well as $k_\mathrm{A}$ and the  reverse intersystem crossing rates $k_\mathrm{RISC}$ and the intersystem crossing rates $k_\mathrm{ISC}$. Transition rates are depicted by open symbols.}
    \label{fig:parameter_ks}
\end{figure}

\begin{figure}[H]
    \centering
    \includegraphics[width=0.5\linewidth]{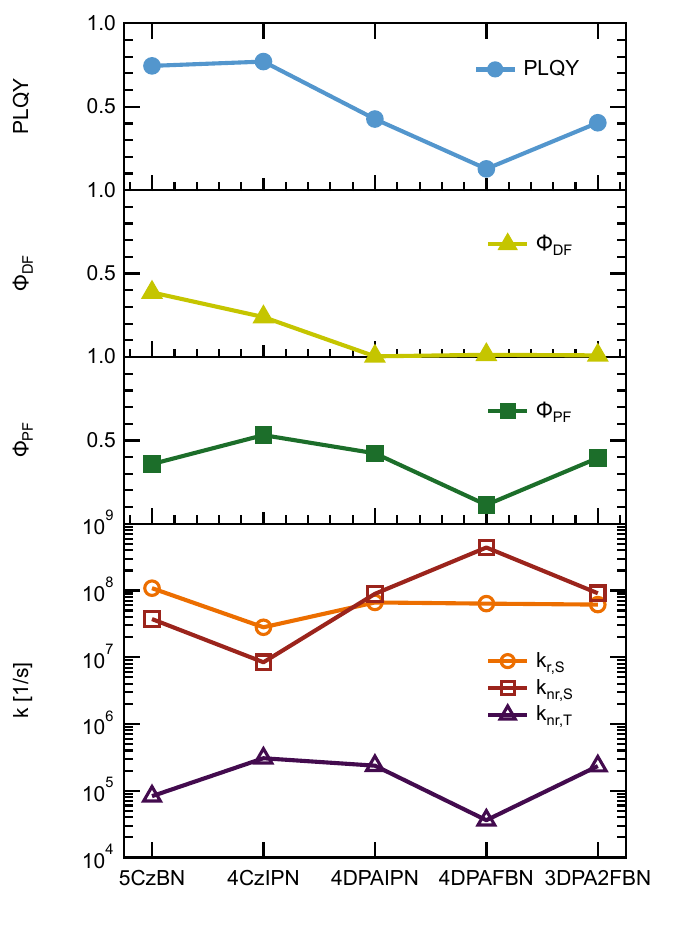}
    \caption{Photoluminescence quantum yields $PLQY$ and efficiencies of delayed $\Phi_{DF}$ and prompt fluorescence efficiency $\Phi_{PF}$ alongside the rates of radiative and non-radiative recombination for singlet ($k_\mathrm{r, S}$, $k_\mathrm{nr,S}$) and triplet states($k_\mathrm{nr,T}$). Transition rates are depicted by open symbols.}
    \label{fig:parameter_PLQYs}
\end{figure}

\section{Laplace Transform of the Gamma Distribution} 

\begin{figure}[H]
    \centering
    \includegraphics[width=0.6\linewidth]{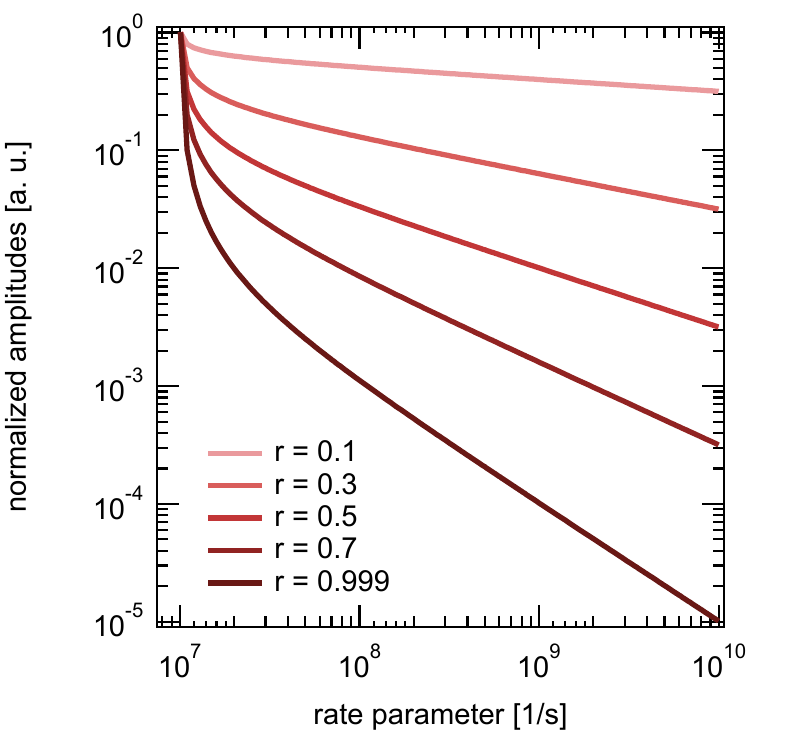}
    \caption{Plot of the inverse Laplace transform of the gamma distribution for different shape parameters $r$.}
    \label{fig:inv_LPT_gamma_r}
\end{figure}

To obtain the distribution of decay rates, a Laplace transform was applied to the gamma distribution in eq (1): 

\begin{equation}
    \mathcal{L}^{-1} [f(t)] = \frac{k^r}{\Gamma(r) \Gamma(1-r)}\cdot\frac{1}{(s-k)^r}
\end{equation}
\\
The resulting rate distribution is shown Figure \ref{fig:inv_LPT_gamma_r} for different values of $r$. There is a sharp increase for rates approaching $k$ and a decrease with $s^{-r}$ for high rates. The distribution resembles a delta peak for $r$ approaching $1$ and becomes broader for $r < 1$.
\noindent The rate distribution for the introduced 'Gamma-Fit' consists of two of these curves and is illustrated using experimental data from 4CzIPN in Figure 3.

\section{Excitation and Photoluminescence spectra of thin film samples}

\begin{figure}[H]
    \centering
    \includegraphics[width=\linewidth]{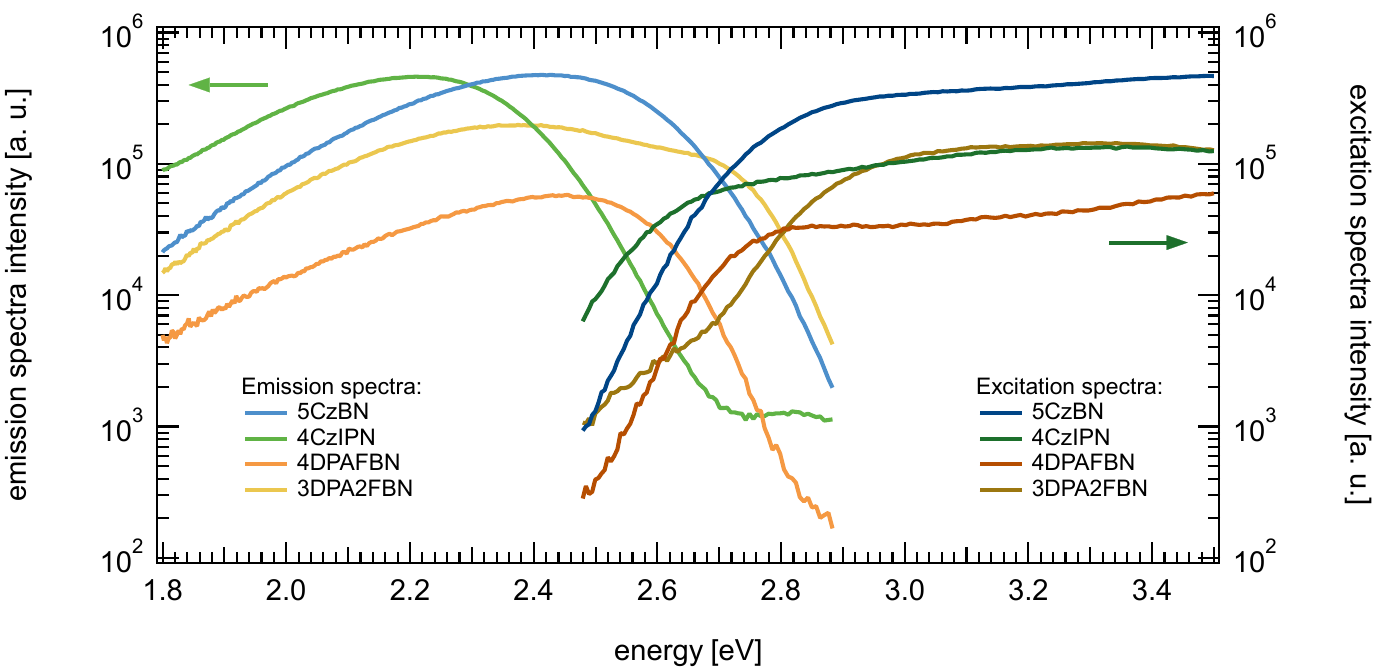}
    \caption{Measured excitation and emission spectra of all TADF emitters.}
    \label{fig:TADF_spectras}
\end{figure}

\section{Computational Data}

\begin{table}[H]
    \centering
    \caption{Optimal $\omega$B97X-D3 range separation parameters obtained for all molecules.}
    \renewcommand{\arraystretch}{1.8}
    \vspace{1em}
    \begin{tabular}{m{5.5em}||m{3.5em}}
        \hline
        Molecule & $\omega$ \\
        \hline
        \hline
        5CzBN & $0.1031$ \\
        \hline
        4CzIPN & $0.1090$ \\
        \hline
        4DPAIPN & $0.1028$ \\
        \hline
        4DPAFBN & $0.0535$ \\
        \hline
        3DPA2FBN & $0.1195$ \\
        \hline
    \end{tabular}
    \label{tab:omega}
\end{table}

\begin{table}[H]
    \centering
    \caption{Computational values of intersystem crossing properties for all molecules, determined with B3LYP-D3/def2-SVPD.}
    \renewcommand{\arraystretch}{1.8}
    \vspace{1em}
    \begin{tabular}{m{5.5em}||m{5.5em}|m{5.5em}|m{5.5em}|m{5.5em}|m{5.5em}|m{5.5em}}
        \hline
        Molecule & $k_\mathrm{ISC}$ $[\mathrm{s^{-1}}]$ & $k_\mathrm{RISC}$ $[\mathrm{s^{-1}}]$ & $\lambda_{T\rightarrow S}$ $[\mathrm{meV}]$ & $E_\mathrm{A}$ $[\mathrm{meV}]$ & $\Delta E_{T\rightarrow S}$ $[\mathrm{meV}]$ & $H^\mathbf{T}_\mathrm{SO}$ $[\mathrm{cm^{-1}}]$ \\
        \hline
        \hline
        5CzBN & $3.48\cdot 10^{7}$ & $6.90\cdot 10^{6}$ & $134$ & $100$ & $97$  & $0.69$ \\
        \hline
        4CzIPN & $1.96\cdot 10^{7}$ & $2.16\cdot 10^{6}$ & $22$  & $75$  & $59$  & $0.15$ \\
        \hline
        4DPAIPN & $6.68\cdot 10^{8}$ & $4.47\cdot 10^{5}$ & $105$ & $207$ & $190$ & $1.33$ \\
        \hline
        4DPAFBN & $3.42\cdot 10^{8}$ & $2.38\cdot 10^{4}$ & $192$ & $271$ & $264$ & $1.23$ \\
        \hline
        3DPA2FBN & $3.16\cdot 10^{7}$ & $2.86\cdot 10^{2}$ & $175$ & $332$ & $307$ & $0.44$ \\
        \hline
    \end{tabular}
    \label{tab:results_B3LYP}
\end{table}

\begin{table}[H]
    \centering
    \caption{Computational values of intersystem crossing properties for all molecules, determined with OT-$\omega$B97X-D3/def2-SVPD.}
    \renewcommand{\arraystretch}{1.8}
    \vspace{1em}
    \begin{tabular}{m{5.5em}||m{5.5em}|m{5.5em}|m{5.5em}|m{5.5em}|m{5.5em}|m{5.5em}}
        \hline
        Molecule & $k_\mathrm{ISC}$ $[\mathrm{s^{-1}}]$ & $k_\mathrm{RISC}$ $[\mathrm{s^{-1}}]$ & $\lambda_{T\rightarrow S}$ $[\mathrm{meV}]$ & $E_\mathrm{A}$ $[\mathrm{meV}]$ & $\Delta E_{T\rightarrow S}$ $[\mathrm{meV}]$ & $H^\mathbf{T}_\mathrm{SO}$ $[\mathrm{cm^{-1}}]$ \\
        \hline
        \hline
        5CzIPN & $6.22\cdot 10^{7}$ & $4.59\cdot 10^{6}$ & $168$ & $100$ & $91$  & $0.60$ \\
        \hline
        4CzIPN & $4.74\cdot 10^{6}$ & $2.76\cdot 10^{6}$ & $47$  & $59$  & $58$  & $0.15$ \\
        \hline
        4DPAIPN & $6.64\cdot 10^{8}$ & $2.87\cdot 10^{3}$ & $94$  & $331$ & $258$ & $1.15$ \\
        \hline
        4DPAFBN & $3.43\cdot 10^{8}$ & $7.93\cdot 10^{5}$ & $280$ & $169$ & $155$ & $1.08$ \\
        \hline
        3DPA2FBN & $4.99\cdot 10^{7}$ & $7.16\cdot 10^{2}$ & $225$ & $295$ & $290$ & $0.36$ \\
        \hline
    \end{tabular}
    \label{tab:results_wB97X}
\end{table}

\section{Statistical Modeling}\label{sec:Stat}

\subsection{Predictor Construction and Data Preparation}\label{sec:Pred}

To investigate the lack of systematic correlation between the values calculated \textit{via} the computational protocol and the experimental benchmark data, we evaluated four potential sources of the observed discrepancies between the experiments and our computational modeling:
\medskip
\begin{enumerate}
    \item[a.] 
    Model inadequacy: The Marcus-like approach may not provide a uniform description across the chemical space studied. This applies to both theory and experiment.
    \item[b.] 
    Quantum chemical limitations: The chosen level of theory may exhibit inconsistent performance depending on the molecular structure.
    \item[c.] 
    Conformational complexity: The computational protocol relies on a static geometry approach, neglecting potential contributions from a broad conformational ensemble to the photophysical properties.
    \item[d.]
    Alternative relaxation pathways: The computational protocol does not take higher-lying triplet state crossing pathways into account, which influence the observed properties.
\end{enumerate}
\medskip
We constructed statistical predictors to test the validity of these hypotheses and quantify their impact on the discrepancy between computational and experimental data.\\
The dataset is comprised of range-normalized absolute errors (RNAEs) of all properties accessible by both experimental measurements and theoretical calculations to guarantee comparability:

\begin{equation}
    RNAE = \frac{|x_\mathrm{comp.}-x_\mathrm{exp.}|}{x_\mathrm{exp.}(max.)-x_\mathrm{exp.}(min.)}
\end{equation}
\\
$x_\mathrm{comp.}$ and $y_\mathrm{exp.}$ are the corresponding pairs of computational and experimental values, respectively. Included properties are the ISC rate $k_\mathrm{ISC}$, the RISC rate $k_\mathrm{ISC}$, the RISC reorganization energy $\lambda_{T\rightarrow S}$, the RISC activation energy $E_\mathrm{A}$, the RISC singlet--triplet energy gap $\Delta E_{T\rightarrow S}$, and the RISC SOC $H^\mathbf{T}_\mathrm{SO}$. Given that $k_\mathrm{ISC}$ and $k_\mathrm{RISC}$ span several orders of magnitude, the logarithmic values were used for calculation of the corresponding RNAEs. All RNAEs for both functionals discussed are given in Tables \ref{tab:RNAE_B3LYP} and \ref{tab:RNAE_wB97X} and are used to gauge limitations of the utilized level of theory.

\begin{table}[H]
    \centering
    \caption{Range-normalized absolute computational modeling errors of all properties accessible by both experimental measurements and theoretical calculations for all molecules using B3LYP.}
    \renewcommand{\arraystretch}{1.8}
    \vspace{1em}
    \begin{tabular}{m{5.5em}||m{5.5em}|m{5.5em}|m{5.5em}|m{5.5em}|m{5.5em}|m{5.5em}}
        \hline
        Molecule & $k_\mathrm{ISC}$ & $k_\mathrm{RISC}$ & $\lambda_{T\rightarrow S}$ & $E_\mathrm{A}$ & $\Delta E_{T\rightarrow S}$ & $H^\mathbf{T}_\mathrm{SO}$ \\
        \hline
        \hline
        5CzBN & $0.31$ & $0.89$ & $0.34$ & $0.18$ & $0.15$ & $0.97$ \\
        \hline
        4CzIPN & $0.04$ & $0.28$ & $0.69$ & $0.05$ & $0.01$ & $0.06$ \\
        \hline
        4DPAIPN & $1.30$ & $0.20$ & $0.16$ & $0.42$ & $0.30$ & $2.63$ \\
        \hline
        4DPAFBN & $0.37$ & $0.17$ & $0.32$ & $0.40$ & $0.40$ & $1.90$ \\
        \hline
        3DPA2FBN & $0.44$ & $2.18$ & $0.92$ & $1.27$ & $1.38$ & $0.52$ \\
        \hline
    \end{tabular}
    \label{tab:RNAE_B3LYP}
\end{table}

\begin{table}[H]
    \centering
    \caption{Range-normalized absolute computational modeling errors of all properties accessible by both experimental measurements and theoretical calculations for all molecules using $\omega$B97X-D3.}
    \renewcommand{\arraystretch}{1.8}
    \vspace{1em}
    \begin{tabular}{m{5.5em}||m{5.5em}|m{5.5em}|m{5.5em}|m{5.5em}|m{5.5em}|m{5.5em}}
        \hline
        Molecule & $k_\mathrm{ISC}$ & $k_\mathrm{RISC}$ & $\lambda_{T\rightarrow S}$ & $E_\mathrm{A}$ & $\Delta E_{T\rightarrow S}$ & $H^\mathbf{T}_\mathrm{SO}$ \\
        \hline
        \hline
        5CzBN & $0.19$ & $0.76$ & $0.13$ & $0.18$ & $0.20$ & $0.73$ \\
        \hline
        4CzIPN & $0.25$ & $0.36$ & $0.53$ & $0.06$ & $0.00$ & $0.05$ \\
        \hline
        4DPAIPN & $1.30$ & $1.43$ & $0.23$ & $1.27$ & $0.76$ & $2.15$ \\
        \hline
        4DPAFBN & $0.37$ & $0.96$ & $0.88$ & $0.30$ & $0.34$ & $1.51$ \\
        \hline
        3DPA2FBN & $0.53$ & $1.88$ & $1.23$ & $1.01$ & $1.26$ & $0.30$ \\
        \hline
    \end{tabular}
    \label{tab:RNAE_wB97X}
\end{table}

\noindent Isolating the specific computational modeling error contribution of the used Marcus-like approach is not possible, as every theory-experiment value pair contributing to the RNAEs in our dataset is based on semiclassical assumptions. In each case, either the computational value, the experimental reference, or both, are derived from Marcus- or Arrhenius-type approaches. However, by ordering the RNAEs according to this scheme, the extent and magnitude of error can at least be estimated for theory and experiment separately. Consequently, Marcus subset $\mathcal{A}$ includes properties where only the experimental values were obtained \textit{via} a Marcus-type framework ($\Delta E_{T\rightarrow S}$, $H^\mathbf{T}_\mathrm{SO}$), Marcus subset $\mathcal{B}$ is the analogue for computational values ($k_\mathrm{ISC}$, $k_\mathrm{RISC}$), and Marcus subset $\mathcal{C}$ contains properties inherent to Marcus theory where all values have been derived from a Marcus-like approach ($\lambda_{T\rightarrow S}$, $E_\mathrm{A}$).\\

\noindent As a predictor for conformational complexity, we chose the Shannon index $H^\prime$, a popular metric for quantifying diversity in ecological contexts\cite{shannon1948mathematical, spellerberg2003tribute}:

\begin{equation}
    H^\prime = - \sum\limits_{i=1}^{N}p_\mathrm{i}\ln{p_\mathrm{i}}
\end{equation}
\\
$N$ is the number of conformers for a given molecule and $p_\mathrm{i}$ are the Boltzmann probabilities of each conformer. Both properties can be obtained from the GOAT conformational sampling runs detailed in the \textbf{Quantum-Chemical Calculations} section of the \textbf{Experimental and Computational Details}. Table \ref{tab:Shannon} contains the calculated Shannon indices for all molecules.
\begin{table}[H]
    \centering
    \caption{Shannon indices for all molecules.}
    \renewcommand{\arraystretch}{1.8}
    \vspace{1em}
    \begin{tabular}{m{5.5em}||m{3.5em}}
        \hline
        Molecule & Shannon index $H^\prime$ \\
        \hline
        \hline
        5CzBN & $0.00$ \\
        \hline
        4CzIPN & $0.43$ \\
        \hline
        4DPAIPN & $0.03$ \\
        \hline
        4DPAFBN & $0.05$ \\
        \hline
        3DPA2FBN & $0.72$ \\
        \hline
    \end{tabular}
    \label{tab:Shannon}
\end{table}
\noindent The likelihood of a crossing transition between two states is mainly governed by their energy gap and their coupling strength. Consequently, we devised a Pathway index $P$ that estimates the favorability of a transition \textit{via} a higher triplet state $T_\mathrm{2}$ or $T_\mathrm{3}$ over a transition \textit{via} the $T_\mathrm{1}$ state from their respective singlet--triplet gaps $\Delta E_{T\rightarrow S}$ and SOC constants $H^\mathbf{T}_\mathrm{SO}$:

\begin{equation}
    P = \frac{(|H^\mathbf{T_\mathrm{3}}_\mathrm{SO}|^2\exp{(-\Delta E_{T_\mathrm{3}\rightarrow S_\mathrm{1}})}) + (|H^\mathbf{T_\mathrm{2}}_\mathrm{SO}|^2\exp{(-\Delta E_{T_\mathrm{2}\rightarrow S_\mathrm{1}})})}{(|H^\mathbf{T_\mathrm{1}}_\mathrm{SO}|^2\exp{(-\Delta E_{T_\mathrm{1}\rightarrow S_\mathrm{1}})})}
\end{equation}
\\
Adiabatic energies were utilized for the $S_\mathrm{1}$ and $T_\mathrm{1}$ states. In absence of optimized geometries for the $T_\mathrm{2}$ and $T_\mathrm{3}$ states, their vertical energies were adjusted using the $T_\mathrm{1}$ adiabatic stabilization energy (defined as the vertical-to-adiabatic energy difference). This approximation is justified by the consistent CT character across all involved triplet states (Table \ref{tab:CT}). SOC constants were evaluated at the $T_\mathrm{1}$ equilibrium geometry, and the resulting Pathway indices for all molecules and functionals are summarized in Table \ref{tab:Pathway}.

\begin{table}[H]
    \centering
    \caption{Charge-transfer character of the $T_\mathrm{1}$, $T_\mathrm{2}$, and $T_\mathrm{3}$ transitions based on the $T_\mathrm{1}$ equilibrium geometry for each molecule.}
    \renewcommand{\arraystretch}{1.8}
    \vspace{1em}
    \begin{tabular}{m{5.5em}||m{5.5em}|m{5.5em}|m{5.5em}||m{5.5em}|m{5.5em}|m{5.5em}}
        \hline
        \multirow{3}{*}{Molecule} & \multicolumn{6}{c}{CT Character} \\
        \hhline{~------}
        & \multicolumn{3}{c||}{B3LYP} & \multicolumn{3}{c}{$\omega$B97X-D} \\
        \hhline{~------}
        & $T_\mathrm{1}$ & $T_\mathrm{2}$ & $T_\mathrm{3}$ & $T_\mathrm{1}$ & $T_\mathrm{2}$ & $T_\mathrm{3}$ \\
        \hline
        5CzBN & $0.639$ & $0.734$ & $0.766$ & $0.634$ & $0.713$ & $0.771$ \\
        \hline
        4CzIPN & $0.800$ & $0.722$ & $0.796$ & $0.784$ & $0.710$ & $0.793$ \\
        \hline
        4DPAIPN & $0.625$ & $0.549$ & $0.710$ & $0.620$ & $0.572$ & $0.709$ \\
        \hline
        4DPAFBN & $0.545$ & $0.579$ & $0.667$ & $0.578$ & $0.595$ & $0.680$ \\
        \hline
        3DPA2FBN & $0.498$ & $0.599$ & $0.691$ & $0.509$ & $0.568$ & $0.694$ \\
        \hline
    \end{tabular}
    \label{tab:CT}
\end{table}

\begin{table}[H]
    \centering
    \caption{Pathway indices for all molecules obtained with B3LYP and $\omega$B97X-D3.}
    \renewcommand{\arraystretch}{1.8}
    \vspace{1em}
    \begin{tabular}{m{5.5em}||m{5.5em}|m{5.5em}}
        \hline
        \multirow{2}{*}{Molecule} & \multicolumn{2}{c}{Pathway index $P$} \\
        \hhline{~--}
         & B3LYP & $\omega$B97X-D3 \\
        \hline
        5CzBN & $0.44$ & $0.55$ \\
        \hline
        4CzIPN & $0.79$ & $0.85$ \\
        \hline
        4DPAIPN & $3.87$ & $4.77$ \\
        \hline
        4DPAFBN & $0.91$ & $0.99$ \\
        \hline
        3DPA2FBN & $1.17$ & $2.16$ \\
        \hline
    \end{tabular}
    \label{tab:Pathway}
\end{table}
\clearpage

\subsection{Multivariate Statistical Modeling I}

The data was fitted using a linear mixed model with random intercepts equal to the number of molecules, the restricted maximum likelihood approach and the following regression equation: \cite{lindstrom1988newton}

\begin{equation}
    \begin{split}
            RNAE \sim Marcus \ subset \ + \ level \ of \ theory \ + \ Shannon \ index \\ + \ Pathway \ index \ + \ level \ of \ theory:Pathway \ index
    \end{split}
    \label{eq:LMMI}
\end{equation}
\\
The last term is an interaction term used to test the influence of the level of theory on the Pathway index. Results are given in Table \ref{tab:LMMIv1}.

\begin{figure}[H]
    \centering
    \includegraphics[width=\linewidth]{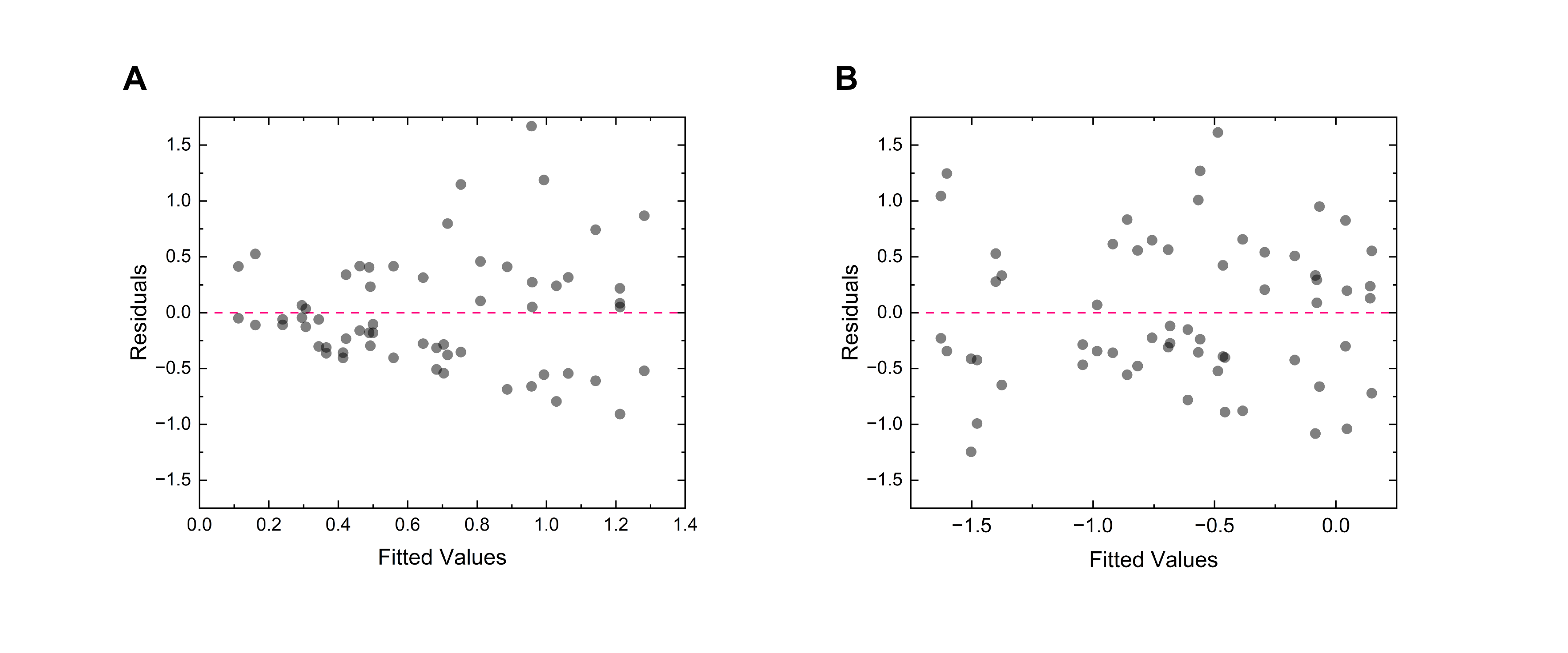}
    \caption{Fitted values plotted against residuals for (A) range-normalized absolute computational modeling errors (RNAEs) and (B) Box-Cox transformed RNAEs.}
    \label{fig:ResFit}
\end{figure}

\begin{figure}[H]
    \centering
    \includegraphics[width=\linewidth]{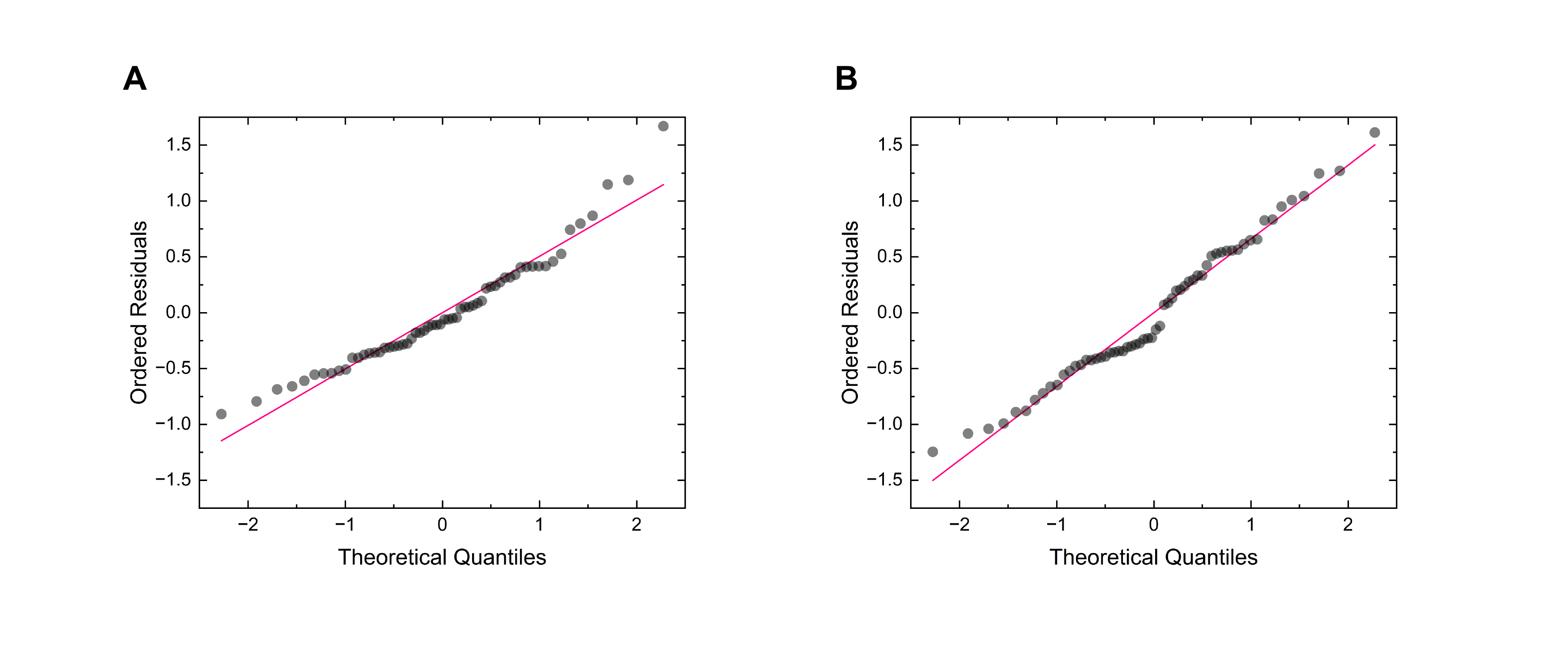}
    \caption{Quantile-quantile plots for (A) range-normalized absolute computational modeling errors (RNAEs) and (B) Box-Cox transformed RNAEs.}
    \label{fig:QQ}
\end{figure}

\begin{table}[H]
    \centering
    \caption{Results of fitting the range-normalized absolute computational modeling errors using eq (\ref{eq:LMMI}).}
    \renewcommand{\arraystretch}{1.8}
    \vspace{1em}
    \begin{tabular}{m{9.5em}||m{5.5em}|m{5.5em}|m{5.5em}|m{5.5em}|m{7.5em}}
        \hline
        Predictor & Coefficient ($\beta$) & Standard Error & Standard Score & $p$-Value & 95\% Confidence Interval \\
        \hline
        \hline
        \multicolumn{6}{l}{Fixed Effects} \\
        \hline
        \hline
        Intercept & $0.249$ & $0.325$ & $0.767$ & $0.443$ & $[-0.387,\,0.8853]$ \\
        \hline
        Level of Theory $\omega$B97X-D3 (Ref.: B3LYP) & $-0.117$ & $0.217$ & $-0.539$ & $0.590$ & $[-0.543,\,0.309]$ \\
        \hline
        Marcus Subset $\mathcal{A}$ (Ref.: Marcus Subset $\mathcal{C}$) & $0.253$ & $0.172$ & $1.474$ & $0.140$ & $[-0.083,\,0.589]$ \\
        \hline
        Marcus Subset $\mathcal{B}$ (Ref.: Marcus Subset $\mathcal{C}$) & $0.183$ & $0.172$ & $1.065$ & $0.287$ & $[-0.154,\,0.519]$ \\
        \hline
        Shannon Index & $0.338$ & $0.552$ & $0.613$ & $0.540$ & $[-0.744,\,1.420]$ \\
        \hline
        Pathway Index & $0.114$ & $0.138$ & $0.824$ & $0.410$ & $[-0.157,\,0.386]$ \\
        \hline
        Level of Theory $\times$ Pathway Index Interaction & $0.071$ & $0.105$ & $0.675$ & $0.499$ & $[-0.135,\,0.277]$ \\
        \hline
        \multicolumn{6}{l}{Random Effects} \\
        \hline
        \hline
        Molecule (Variance) & $0.095$ &  &  &  &  \\
        \hline
        Residual (Scale) & $0.224$ &  &  &  &  \\
        \hline
    \end{tabular}
    \label{tab:LMMIv1}
\end{table}
\noindent To check model validity, we tested homoscedasticity and normality of residuals, the corresponding plots are show in Figure \ref{fig:ResFit}A and Figure \ref{fig:QQ}A, respectively. To ensure model stability, multicollinearity was assessed using generalized variance inflation factors (see Table \ref{tab:LMMIv1_GVIF}).

\begin{table}[H]
    \centering
    \caption{Generalized variance inflation factors (GVIFs) and scaled GVIFs for all predictor variables in Table \ref{tab:LMMIv1}.}
    \renewcommand{\arraystretch}{1.8}
    \vspace{1em}
    \begin{tabular}{m{9.5em}||m{5.5em}|m{3.5em}|m{5.5em}}
        \hline
        Variable & GVIF & DF & GVIF$^\mathrm{\frac{1}{2DF}}$ \\
        \hline
        \hline
        Level of Theory & $2.3987$ & $1$ & $1.549$ \\
        \hline
        Marcus Subset & $1.000$ & $2$ & $1.000$ \\
        \hline
        Shannon Index & $1.033$ & $1$ & $1.017$ \\
        \hline
        Pathway Index & $2.707$ & $1$ & $1.645$ \\
        \hline
        Level of Theory $\times$ Pathway Index Interaction & $4.461$ & $1$ & $2.112$ \\
        \hline
    \end{tabular}
    \label{tab:LMMIv1_GVIF}
\end{table}
\noindent As the uneven distribution in Figure \ref{fig:ResFit}A indicates heteroscedasticity, the RNAEs were transformed using a Box-Cox transformation to stabilize variance. \cite{box1964analysis} Furthermore, minor structural multicollinearity was observed for the level of theory and for its interaction with the Pathway index. This can be addressed by mean-centering the Pathway index, although the observed collinearity between the level of theory and the Pathway index is a direct consequence of the electronic properties being functional-dependent, so the two will be naturally connected. The data was then re-evaluated using the transformed values. Results are given in Table \ref{tab:LMMIv2}. New validity checks are given in Figure \ref{fig:ResFit}B, Figure \ref{fig:QQ}B, and Table \ref{tab:LMMIv2_GVIF}.

\begin{table}[H]
    \centering
    \caption{Results of fitting the Box-Cox transformed range-normalized absolute computational modeling errors using eq (\ref{eq:LMMI}).}
    \renewcommand{\arraystretch}{1.8}
    \vspace{1em}
    \begin{tabular}{m{9.5em}||m{5.5em}|m{5.5em}|m{5.5em}|m{5.5em}|m{7.5em}}
        \hline
        Predictor & Coefficient ($\beta$) & Standard Error & Standard Score & $p$-Value & 95\% Confidence Interval \\
        \hline
        \hline
        \multicolumn{6}{l}{Fixed Effects} \\
        \hline
        \hline
        Intercept & $-0.883$ & $0.437$ & $-2.024$ & $0.043$ & $[-1.739,\,-0.028]$ \\
        \hline
        Level of Theory $\omega$B97X-D3 (Ref.: B3LYP) & $0.066$ & $0.205$ & $0.323$ & $0.747$ & $[-0.336,\,0.468]$ \\
        \hline
        Marcus Subset $\mathcal{A}$ (Ref.: Marcus Subset $\mathcal{C}$) & $0.124$ & $0.222$ & $0.559$ & $0.576$ & $[-0.311,\,0.559]$ \\
        \hline
        Marcus Subset $\mathcal{B}$ (Ref.: Marcus Subset $\mathcal{C}$) & $0.226$ & $0.222$ & $1.018$ & $0.309$ & $[-0.209,\,0.661]$ \\
        \hline
        Shannon Index & $0.275$ & $1.097$ & $0.250$ & $0.802$ & $[-1.876,\,2.426]$ \\
        \hline
        Pathway Index (Centered) & $0.088$ & $0.260$ & $0.338$ & $0.735$ & $[-0.421,\,0.597]$ \\
        \hline
        Level of Theory $\times$ Pathway Index Interaction & $0.122$ & $0.141$ & $0.863$ & $0.388$ & $[-0.155,\,0.398]$ \\
        \hline
        \multicolumn{6}{l}{Random Effects} \\
        \hline
        \hline
        Molecule (Variance) & $0.433$ &  &  &  &  \\
        \hline
        Residual (Scale) & $0.686$ &  &  &  &  \\
        \hline
    \end{tabular}
    \label{tab:LMMIv2}
\end{table}

\begin{table}[H]
    \centering
    \caption{Generalized variance inflation factors (GVIFs) and scaled GVIFs for all predictor variables in Table \ref{tab:LMMIv2}.}
    \renewcommand{\arraystretch}{1.8}
    \vspace{1em}
    \begin{tabular}{m{9.5em}||m{5.5em}|m{3.5em}|m{5.5em}}
        \hline
        Variable & GVIF & DF & GVIF$^\mathrm{\frac{1}{2DF}}$ \\
        \hline
        \hline
        Level of Theory & $1.025$ & $1$ & $1.012$ \\
        \hline
        Marcus Subset & $1.000$ & $2$ & $1.000$ \\
        \hline
        Shannon Index & $1.034$ & $1$ & $1.017$ \\
        \hline
        Pathway Index (Centered) & $2.707$ & $1$ & $1.645$ \\
        \hline
        Level of Theory $\times$ Pathway Index Interaction & $2.619$ & $1$ & $1.618$ \\
        \hline
    \end{tabular}
    \label{tab:LMMIv2_GVIF}
\end{table}
\noindent From Table \ref{tab:LMMIv2}, no single predictor could be identified as a statistically significant source of computational modeling error. The following section provides supplemental analyses and a re-evaluation of the model leading to statistically significant effects.

\subsection{Further Statistical Analyses}

Comparison of marginal ($R^2 = 0.074$) and conditional ($R^2 = 0.507$) coefficients of determination suggests that the primary source of variance resides in inherent molecular differences.\\
This is further corroborated by shrinkage analysis of the random effects. While the random intercept for 4CzIPN indicates a significantly lower computational modeling error than predictors would suggest, all other molecules exhibit varying higher baseline errors. All random intercepts can be found in Table \ref{tab:random_effects}.

\begin{table}[H]
    \centering
    \caption{Individual random effects of the linear mixed model in Table \ref{tab:LMMIv2}.}
    \renewcommand{\arraystretch}{1.8}
    \vspace{1em}
    \begin{tabular}{m{5.5em}||m{5.5em}}
        \hline
        Molecule & Random Intercept \\
        \hline
        \hline
        5CzBN & $0.434$ \\
        \hline
        4CzIPN & $-0.761$ \\
        \hline
        4DPAIPN & $0.251$ \\
        \hline
        4DPAFBN & $0.069$ \\
        \hline
        3DPA2FBN & $0.006$ \\
        \hline
    \end{tabular}
    \label{tab:random_effects}
\end{table}
\noindent This uneven distribution explains the lack of predictive power of the model, at least for this constellation of predictors, as the estimates for the distinct molecules are pulled toward a mean that is ultimately unrepresentative of the wider ensemble.\\ 
To ensure that these molecular differences do not introduce inconsistencies that affect our evaluation of theoretical accuracy, we treated the molecule grouping as a fixed effect and fitted the data using ordinary least squares regression with 4CzIPN as the reference and the following regression equation:

\begin{equation}
    RNAE(Box-Cox) \sim Marcus \ subset \ + \ level \ of \ theory \cdot C(Molecule)
    \label{eq:OLSI}
\end{equation}
\\
The results can be found in Table \ref{tab:OLSI}.

\begin{table}[H]
    \centering
    \caption{Results of fitting the Box-Cox transformed range-normalized absolute computational modeling errors using eq (\ref{eq:OLSI}). Statistical significance is indicated by an asterisk.}
    \renewcommand{\arraystretch}{1.8}
    \vspace{1em}
    \begin{tabular}{m{9.5em}||m{5.5em}|m{5.5em}|m{5.5em}|m{5.5em}|m{7.5em}}
        \hline
        Predictor & Coefficient ($\beta$) & Standard Error & t-Statistic & $p$-Value & 95\% Confidence Interval \\
        \hline
        \hline
        Intercept (Ref: 4CzIPN) & $-1.740$ & $0.319$ & $-5.450$ & $<0.001^*$ & $[-2.382,\, -1.098]$ \\
        \hline
        \multicolumn{6}{l}{Molecule Effects (Ref.: 4CzIPN)} \\
        \hline
        \hline
        3DPA2FBN & $1.633$ & $0.412$ & $3.692$ & $<0.001^*$ & $[0.804,\,2.462]$ \\
        \hline
        4DPAFBN & $0.948$ & $0.412$ & $2.300$ & $0.026^*$ & $[0.119,\,1.777]$ \\
        \hline
        4DPAIPN & $1.119$ & $0.412$ & $2.715$ & $0.009^*$ & $[0.290,\,1.948]$ \\
        \hline
        5CzBN & $0.828$ & $0.412$ & $2.008$ & $0.050^*$ & $[-0.001,\,1.657]$ \\
        \hline
        \multicolumn{6}{l}{Interaction Terms (Ref.: 4CzIPN)} \\
        \hline
        \hline
        Level of Theory $\times$ 3DPA2FBN & $-0.183$ & $0.583$ & $-0.315$ & $0.754$ & $[-1.356,\,0.988]$ \\
        \hline
        Level of Theory $\times$ 4DPAFBN & $0.150$ & $0.583$ & $0.257$ & $0.798$ & $[-1.022,\,1.322]$ \\
        \hline
        Level of Theory $\times$ 4DPAIPN & $0.452$ & $0.583$ & $0.776$ & $0.441$ & $[-0.719,\,1.625]$ \\
        \hline
        Level of Theory $\times$ 5CzBN & $-0.308$ & $0.583$ & $-0.528$ & $0.600$ & $[-1.480,\,0.864]$ \\
        \hline
        \multicolumn{6}{l}{Model Summary} \\
        \hline
        \hline
        R$^2$ (Adj. R$^2$) & \multicolumn{5}{l}{$0.445\,(0.318)$} \\
        \hline
        F-Statistic & $3.502$ &  &  &  $0.001$ &  \\
        \hline
    \end{tabular}
    \label{tab:OLSI}
\end{table}
\noindent All respective interaction terms lack statistical significance. This indicates a consistent behavior across functionals, meaning that switching from B3LYP to $\omega$B97X-D3 does not shift the computational modeling error profile of the other molecules in a manner significantly different from the reference. Notably, the main effects for each molecule were both positive and highly significant, with coefficients increasing in magnitude from 5CzBN \textless \ 4DPAFBN \textless \ 4DPAIPN \textless \ 3DPA2FBN. This hierarchy confirms a stepwise increase in computational modeling difficulty that correlates with conformational diversity.\\
To assess the sensitivity of the Pathway and Shannon indices, a leave-one-out cross-validation (LOOCV) was performed by fitting the data using ordinary least squares regression while excluding one molecule at a time. The following regression equation was used:

\begin{equation}
    RNAE(Box-Cox) \sim Shannon \ index \ + \ Pathway \ index
\end{equation}
\\
The results are shown in Table \ref{tab:LOOCV}. The Pathway index emerged as a robust and consistently significant predictor ($p < 0.05$ across all iterations), despite maintaining a smaller absolute effect size than the Shannon index. While the Shannon index yielded larger coefficients, it failed to reach statistical significance in the majority of cases and exhibited high volatility, with the sign of the coefficient ($\beta$) fluctuating between positive and negative values. This instability suggests the Shannon index is an insufficient proxy for the complexities of conformational diversity. Instead, the variance is more reliably captured by the measure of alternative relaxation pathways. Physically, conformational diversity in these systems is governed by the structure of the donor subunits and their accessible degrees of freedom. The transition from rigid, planar carbazole (Cz) to flexible diphenylamine (DPA) introduces low-frequency torsional modes that are easily populated at room temperature, given no steric hindrance by neighboring substituents.

\begin{table}[H]
    \centering
    \caption{$p$-Values and Coefficients of Shannon index and Pathway index from the leave-one-out cross-validation. Statistical significance is indicated by an asterisk.}
    \renewcommand{\arraystretch}{1.8}
    \vspace{1em}
    \begin{tabular}{m{5.5em}||m{5.5em}|m{5.5em}||m{5.5em}|m{5.5em}}
        \hline
        \multirow{2}{=}{Excluded Molecule} & \multicolumn{2}{c||}{Shannon Index} & \multicolumn{2}{c}{Pathway Index} \\
        \hhline{~----}
         & $p$-Value & Coefficient ($\beta$) & $p$-Value & Coefficient ($\beta$) \\
        \hline
        \hline
        5CzBN & $0.274$ & $0.526$ & $0.007^*$ & $0.264$ \\
        \hline
        4CzIPN & $0.013^*$ & $0.832$ & $0.024^*$ & $0.184$ \\
        \hline
        4DPAIPN & $0.286$ & $-0.589$ & $0.006^*$ & $0.938$ \\
        \hline
        4DPAFBN & $0.126$ & $0.648$ & $0.001^*$ & $0.276$ \\
        \hline
        3DPA2FBN & $0.002^*$ & $-1.991$ & $0.021^*$ & $0.164$ \\
        \hline
    \end{tabular}
    \label{tab:LOOCV}
\end{table}

\subsection{Multivariate Statistical Modeling II}

\begin{figure}[H]
    \centering
    \includegraphics[width=\linewidth]{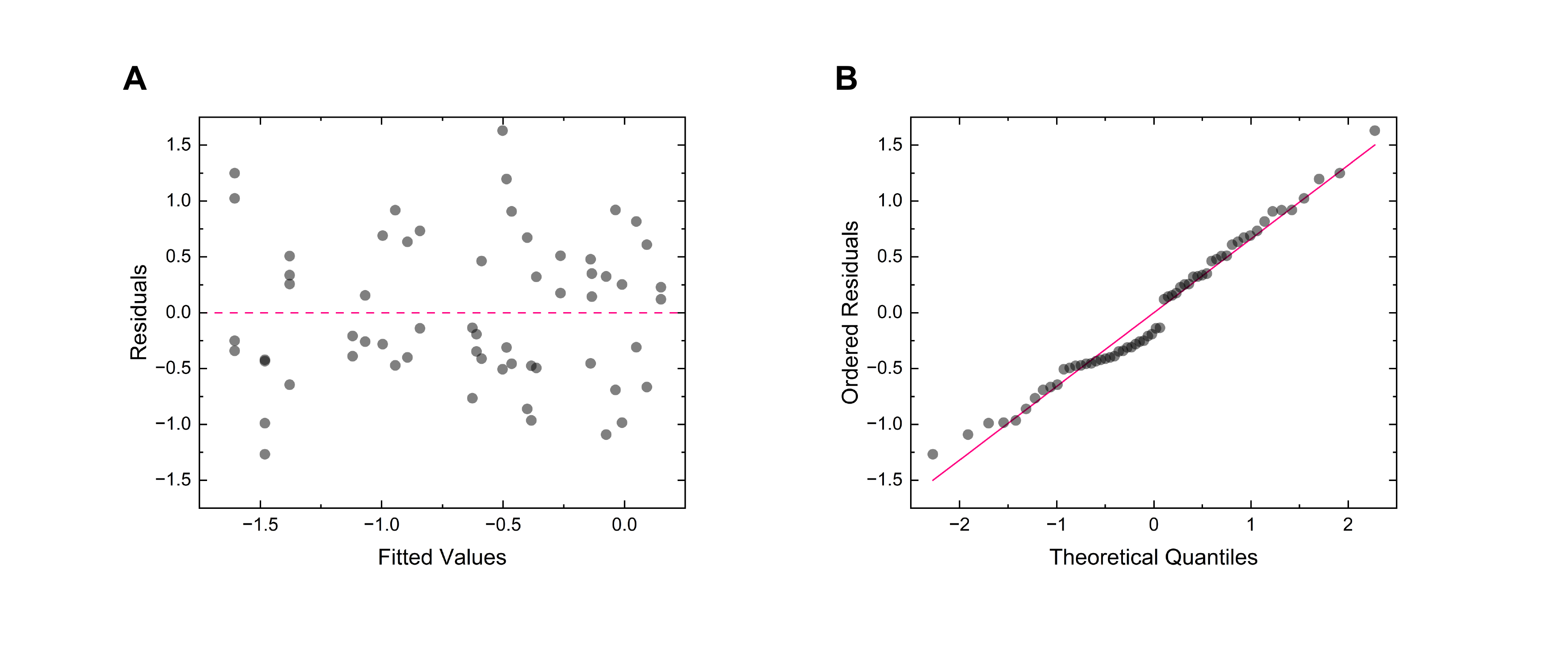}
    \caption{(A) Fitted values plotted against residuals and (B) quantile-quantile plots for Box-Cox transformed range-normalized absolute computational modeling errors.}
    \label{fig:LMMII}
\end{figure}

In order to assess the underlying driver of conformational diversity as explained in the previous section, we exchanged the continuous Shannon index with a categorical Substituent index that identifies each molecule according to its donor subunits (Cz or DPA). Consequently, the new regression equation becomes:

\begin{equation}
    \begin{split}
        RNAE(Box-Cox) \sim Marcus \ subset \ + \ level \ of \ theory \ + \ Substituent \ index \\ + \ Pathway \ index \ (centered) \ + \ level \ of \ theory:Pathway \ index
    \end{split}
    \label{eq:LMMII}
\end{equation}
\\
The model was fitted in the same way as described above, including Box-Cox transformation of the RNAEs and mean-centering of the Pathway index. The results are given in Table \ref{tab:LMMII}, with the necessary checks for homoscedasticity, normality of residuals, and multicollinearity in Figure \ref{fig:LMMII} and Table \ref{tab:LMMII_GVIF}, respectively.
\begin{table}[H]
    \centering
    \caption{Results of fitting the Box-Cox transformed range-normalized absolute computational modeling errors using eq (\ref{eq:LMMII}). Statistical significance is indicated by an asterisk.}
    \label{tab:LMMII}
    \renewcommand{\arraystretch}{1.8}
    \vspace{1em}
    \begin{tabular}{m{9.5em}||m{5.5em}|m{5.5em}|m{5.5em}|m{5.5em}|m{7.5em}}
        \hline
        Predictor & Coefficient ($\beta$) & Standard Error & Standard Score & $p$-Value & 95\% Confidence Interval \\
        \hline
        \hline
        \multicolumn{6}{l}{Fixed Effects} \\
        \hline
        \hline
        Intercept & $-1.423$ & $0.382$ & $-3.722$ & $<0.001^*$ & $[-2.173,\, -0.674]$ \\
        \hline
        Level of Theory $\omega$B97X-D3 (Ref.: B3LYP) & $0.132$ & $0.193$ & $0.682$ & $0.495$ & $[-0.247,\,0.510]$ \\
        \hline
        Marcus Subset $\mathcal{A}$ (Ref.: Marcus Subset $\mathcal{C}$) & $0.124$ & $0.221$ & $0.561$ & $0.575$ & $[-0.309,\,0.557]$ \\
        \hline
        Marcus Subset $\mathcal{B}$ (Ref.: Marcus Subset $\mathcal{C}$) & $0.226$ & $0.221$ & $1.022$ & $0.307$ & $[-0.207,\,0.659]$ \\
        \hline
        Substituent Index DPA (Ref.: Cz) & $0.952$ & $0.456$ & $2.086$ & $0.037^*$ & $[0.58,\,1.846]$ \\
        \hline
        Pathway Index (Centered) & $-0.084$ & $0.192$ & $-0.437$ & $0.662$ & $[-0.460,\,0.292]$ \\
        \hline
        Level of Theory $\times$ Pathway Index Interaction & $0.158$ & $0.136$ & $1.167$ & $0.243$ & $[-0.108,\,0.424]$ \\
        \hline
        \multicolumn{6}{l}{Random Effects} \\
        \hline
        \hline
        Molecule (Variance) & $0.127$ &  &  &  &  \\
        \hline
        Residual (Scale) & $0.241$ &  &  &  &  \\
        \hline
    \end{tabular}
\end{table}

\begin{table}[H]
    \centering
    \caption{Generalized variance inflation factors (GVIFs) and scaled GVIFs for all predictor variables in Table \ref{tab:LMMII}.}
    \label{tab:LMMII_GVIF}
    \renewcommand{\arraystretch}{1.8}
    \vspace{1em}
    \begin{tabular}{m{9.5em}||m{5.5em}|m{3.5em}|m{5.5em}}
        \hline
        Variable & GVIF & DF & GVIF$^\mathrm{\frac{1}{2DF}}$ \\
        \hline
        \hline
        Level of Theory & $1.036$ & $1$ & $1.018$ \\
        \hline
        Marcus Subset & $1.000$ & $2$ & $1.000$ \\
        \hline
        Shannon Index & $1.500$ & $1$ & $1.224$ \\
        \hline
        Pathway Index (Centered) & $3.206$ & $1$ & $1.790$ \\
        \hline
        Level of Theory $\times$ Pathway Index Interaction & $2.596$ & $1$ & $1.611$ \\
        \hline
    \end{tabular}
\end{table}
\noindent The revised model identifies the substituent index as the primary predictor of computational modeling error. Conversely, computational modeling choices such as the level of theory or the use of a Marcus-like approach were not associated with significant error.\\
The transition from a Cz-based to a DPA-based emitter raises the computational modeling error by close to one unit on the Box-Cox scale. Although the Pathway index does not show statistical significance in this model iteration, a second LOOCV analysis reveals that the improvement in description provided by exchanging the Shannon index for the Substituent index sufficiently explains computational modeling error, rendering alternative relaxation pathways a secondary effect compared to structural flexibility (Table \ref{tab:LOOCVII}).

\begin{table}[H]
    \centering
    \caption{$p$-Values and Coefficients of Substituent index and Pathway index from the leave-one-out cross-validation. Statistical significance is indicated by an asterisk.}
    \renewcommand{\arraystretch}{1.8}
    \vspace{1em}
    \begin{tabular}{m{5.5em}||m{5.5em}|m{5.5em}||m{5.5em}|m{5.5em}}
        \hline
        \multirow{2}{=}{Excluded Molecule} & \multicolumn{2}{c||}{Shannon Index} & \multicolumn{2}{c}{Pathway Index} \\
        \hhline{~----}
         & $p$-Value & Coefficient ($\beta$) & $p$-Value & Coefficient ($\beta$) \\
        \hline
        \hline
        5CzBN & $<0.001^*$ & $1.209$ & $0.443$ & $0.063$ \\
        \hline
        4CzIPN & $0.060$ & $0.517$ & $0.423$ & $0.062$ \\
        \hline
        4DPAIPN & $0.002^*$ & $0.878$ & $0.707$ & $0.103$ \\
        \hline
        4DPAFBN & $<0.001^*$ & $1.260$ & $0.537$ & $-0.066$ \\
        \hline
        3DPA2FBN & $0.022^*$ & $0.660$ & $0.293$ & $0.094$ \\
        \hline
    \end{tabular}
    \label{tab:LOOCVII}
\end{table}
\clearpage

\printbibliography

@Article{Haase2018,
  author    = {Haase, Nils and Danos, Andrew and Pflumm, Christof and Morherr, Antonia and Stachelek, Patrycja and Mekic, Amel and Brütting, Wolfgang and Monkman, Andrew P.},
  journal   = {The Journal of Physical Chemistry C},
  title     = {Kinetic Modeling of Transient Photoluminescence from Thermally Activated Delayed Fluorescence},
  year      = {2018},
  issn      = {1932-7455},
  month     = dec,
  number    = {51},
  pages     = {29173--29179},
  volume    = {122},
  doi       = {10.1021/acs.jpcc.8b11020},
  publisher = {American Chemical Society (ACS)},
}

@Article{Aizawa2020,
  author    = {Aizawa, Naoya and Harabuchi, Yu and Maeda, Satoshi and Pu, Yong-Jin},
  journal   = {Nature Communications},
  title     = {Kinetic prediction of reverse intersystem crossing in organic donor–acceptor molecules},
  year      = {2020},
  issn      = {2041-1723},
  month     = aug,
  number    = {1},
  pages     = {3909},
  volume    = {11},
  doi       = {10.1038/s41467-020-17777-2},
  publisher = {Springer Science and Business Media LLC},
}

@Article{Chen2023,
  author    = {Chen, Gary and Swartzfager, John R. and Asbury, John B.},
  journal   = {Journal of the American Chemical Society},
  title     = {Matrix Dynamics and Their Crucial Role in Non-radiative Decay during Thermally Activated Delayed Fluorescence},
  year      = {2023},
  issn      = {1520-5126},
  month     = nov,
  number    = {46},
  pages     = {25495--25504},
  volume    = {145},
  doi       = {10.1021/jacs.3c11719},
  publisher = {American Chemical Society (ACS)},
}

@Article{Streiter2020,
  author    = {Streiter, Martin and Fischer, Tillmann G. and Wiebeler, Christian and Reichert, Sebastian and Langenickel, Jörn and Zeitler, Kirsten and Deibel, Carsten},
  journal   = {The Journal of Physical Chemistry C},
  title     = {Impact of Chlorine on the Internal Transition Rates and Excited States of the Thermally Delayed Activated Fluorescence Molecule 3CzClIPN},
  year      = {2020},
  issn      = {1932-7455},
  month     = jun,
  number    = {28},
  pages     = {15007--15014},
  volume    = {124},
  doi       = {10.1021/acs.jpcc.0c03341},
  publisher = {American Chemical Society (ACS)},
}

@Article{Kelly2022,
  author    = {Kelly, Daniel and Franca, Larissa G. and Stavrou, Kleitos and Danos, Andrew and Monkman, Andrew P.},
  journal   = {The Journal of Physical Chemistry Letters},
  title     = {Laplace Transform Fitting as a Tool To Uncover Distributions of Reverse Intersystem Crossing Rates in TADF Systems},
  year      = {2022},
  issn      = {1948-7185},
  month     = jul,
  number    = {30},
  pages     = {6981--6986},
  volume    = {13},
  doi       = {10.1021/acs.jpclett.2c01864},
  publisher = {American Chemical Society (ACS)},
}

@Article{Tsuchiya2021,
  author    = {Tsuchiya, Youichi and Diesing, Stefan and Bencheikh, Fatima and Wada, Yoshimasa and dos Santos, Paloma L. and Kaji, Hironori and Zysman-Colman, Eli and Samuel, Ifor D. W. and Adachi, Chihaya},
  journal   = {The Journal of Physical Chemistry A},
  title     = {Exact Solution of Kinetic Analysis for Thermally Activated Delayed Fluorescence Materials},
  year      = {2021},
  issn      = {1520-5215},
  month     = sep,
  number    = {36},
  pages     = {8074--8089},
  volume    = {125},
  doi       = {https://doi.org/10.1021/acs.jpca.1c04056},
  publisher = {American Chemical Society (ACS)},
}

@Article{Mello1997,
  author    = {de Mello, John C. and Wittmann, H. Felix and Friend, Richard H.},
  journal   = {Advanced Materials},
  title     = {An improved experimental determination of external photoluminescence quantum efficiency},
  year      = {1997},
  issn      = {1521-4095},
  month     = mar,
  number    = {3},
  pages     = {230--232},
  volume    = {9},
  doi       = {10.1002/adma.19970090308},
  publisher = {Wiley},
}

@Article{Stavrou2020,
  author    = {Stavrou, Kleitos and Franca, Larissa G. and Monkman, Andrew P.},
  journal   = {ACS Applied Electronic Materials},
  title     = {Photophysics of TADF Guest–Host Systems: Introducing the Idea of Hosting Potential},
  year      = {2020},
  issn      = {2637-6113},
  month     = aug,
  number    = {9},
  pages     = {2868--2881},
  volume    = {2},
  doi       = {10.1021/acsaelm.0c00514},
  publisher = {American Chemical Society (ACS)},
}

@Article{Silva2019,
  author    = {de Silva, Piotr and Kim, Changhae Andrew and Zhu, Tianyu and Van Voorhis, Troy},
  journal   = {Chemistry of Materials},
  title     = {Extracting Design Principles for Efficient Thermally Activated Delayed Fluorescence (TADF) from a Simple Four-State Model},
  year      = {2019},
  issn      = {1520-5002},
  month     = jun,
  number    = {17},
  pages     = {6995--7006},
  volume    = {31},
  doi       = {10.1021/acs.chemmater.9b01601},
  publisher = {American Chemical Society (ACS)},
}

@Article{Noda2018,
  author    = {Noda, Hiroki and Nakanotani, Hajime and Adachi, Chihaya},
  journal   = {Science Advances},
  title     = {Excited state engineering for efficient reverse intersystem crossing},
  year      = {2018},
  issn      = {2375-2548},
  month     = jun,
  number    = {6},
  volume    = {4},
  pages = {eaao6910},
  doi       = {10.1126/sciadv.aao6910},
  publisher = {American Association for the Advancement of Science (AAAS)},
}

@Article{Im2017,
  author    = {Im, Yirang and Kim, Mounggon and Cho, Yong Joo and Seo, Jeong-A and Yook, Kyoung Soo and Lee, Jun Yeob},
  journal   = {Chemistry of Materials},
  title     = {Molecular Design Strategy of Organic Thermally Activated Delayed Fluorescence Emitters},
  year      = {2017},
  issn      = {1520-5002},
  month     = feb,
  number    = {5},
  pages     = {1946--1963},
  volume    = {29},
  doi       = {10.1021/acs.chemmater.6b05324},
  publisher = {American Chemical Society (ACS)},
}

@Article{Shi2022,
  author    = {Shi, Yuhao and Ma, Huili and Sun, Zhiyu and Zhao, Weijun and Sun, Guangyan and Peng, Qian},
  journal   = {Angewandte Chemie International Edition},
  title     = {Optimal Dihedral Angle in Twisted Donor-Acceptor Organic Emitters for Maximized Thermally Activated Delayed Fluorescence},
  year      = {2022},
  issn      = {1521-3773},
  month     = nov,
  number    = {51},
  volume    = {61},
  pages = {e202213463},
  doi       = {10.1002/anie.202213463},
  publisher = {Wiley},
}

@Article{Huang2018,
  author    = {Huang, Wenliang and Einzinger, Markus and Zhu, Tianyu and Chae, Hyun Sik and Jeon, Soonok and Ihn, Soo-Ghang and Sim, Myungsun and Kim, Sunghan and Su, Mingjuan and Teverovskiy, Georgiy and Wu, Tony and Van Voorhis, Troy and Swager, Timothy M. and Baldo, Marc A. and Buchwald, Stephen L.},
  journal   = {Chemistry of Materials},
  title     = {Molecular Design of Deep Blue Thermally Activated Delayed Fluorescence Materials Employing a Homoconjugative Triptycene Scaffold and Dihedral Angle Tuning},
  year      = {2018},
  issn      = {1520-5002},
  month     = feb,
  number    = {5},
  pages     = {1462--1466},
  volume    = {30},
  doi       = {10.1021/acs.chemmater.7b03490},
  publisher = {American Chemical Society (ACS)},
}

@Article{Stachelek2019,
  author    = {Stachelek, Patrycja and Ward, Jonathan S. and dos Santos, Paloma L. and Danos, Andrew and Colella, Marco and Haase, Nils and Raynes, Samuel J. and Batsanov, Andrei S. and Bryce, Martin R. and Monkman, Andrew P.},
  journal   = {ACS Applied Materials \& Interfaces},
  title     = {Molecular Design Strategies for Color Tuning of Blue TADF Emitters},
  year      = {2019},
  issn      = {1944-8252},
  month     = jul,
  number    = {30},
  pages     = {27125--27133},
  volume    = {11},
  doi       = {10.1021/acsami.9b06364},
  publisher = {American Chemical Society (ACS)},
}

@Article{Weissenseel2019,
  author    = {Weissenseel, Sebastian and Drigo, Nikita A. and Kudriashova, Liudmila G. and Schmid, Markus and Morgenstern, Thomas and Lin, Kun-Han and Prlj, Antonio and Corminboeuf, Clémence and Sperlich, Andreas and Brütting, Wolfgang and Nazeeruddin, Mohammad Khaja and Dyakonov, Vladimir},
  journal   = {The Journal of Physical Chemistry C},
  title     = {Getting the Right Twist: Influence of Donor–Acceptor Dihedral Angle on Exciton Kinetics and Singlet–Triplet Gap in Deep Blue Thermally Activated Delayed Fluorescence Emitter},
  year      = {2019},
  issn      = {1932-7455},
  month     = oct,
  number    = {45},
  pages     = {27778--27784},
  volume    = {123},
  doi       = {10.1021/acs.jpcc.9b08269},
  publisher = {American Chemical Society (ACS)},
}

@Article{Uoyama2012,
  author    = {Uoyama, Hiroki and Goushi, Kenichi and Shizu, Katsuyuki and Nomura, Hiroko and Adachi, Chihaya},
  journal   = {Nature},
  title     = {Highly efficient organic light-emitting diodes from delayed fluorescence},
  year      = {2012},
  issn      = {1476-4687},
  month     = dec,
  number    = {7428},
  pages     = {234--238},
  volume    = {492},
  doi       = {10.1038/nature11687},
  publisher = {Springer Science and Business Media LLC},
}

@Article{Niwa2014,
  author    = {Niwa, Akitsugu and Kobayashi, Takashi and Nagase, Takashi and Goushi, Kenichi and Adachi, Chihaya and Naito, Hiroyoshi},
  journal   = {Applied Physics Letters},
  title     = {Temperature dependence of photoluminescence properties in a thermally activated delayed fluorescence emitter},
  year      = {2014},
  issn      = {1077-3118},
  month     = may,
  number    = {21},
  volume    = {104},
  pages     = {213303},
  doi       = {10.1063/1.4878397},
  publisher = {AIP Publishing},
}

@Article{Ishimatsu2013,
  author    = {Ishimatsu, Ryoichi and Matsunami, Shigeyuki and Shizu, Katsuyuki and Adachi, Chihaya and Nakano, Koji and Imato, Toshihiko},
  journal   = {The Journal of Physical Chemistry A},
  title     = {Solvent Effect on Thermally Activated Delayed Fluorescence by 1,2,3,5-Tetrakis(carbazol-9-yl)-4,6-dicyanobenzene},
  year      = {2013},
  issn      = {1520-5215},
  month     = jun,
  number    = {27},
  pages     = {5607--5612},
  volume    = {117},
  doi       = {10.1021/jp404120s},
  publisher = {American Chemical Society (ACS)},
}

@Article{Dias2017,
  author    = {Dias, Fernando B and Penfold, Thomas J and Monkman, Andrew P},
  journal   = {Methods and Applications in Fluorescence},
  title     = {Photophysics of thermally activated delayed fluorescence molecules},
  year      = {2017},
  issn      = {2050-6120},
  month     = mar,
  number    = {1},
  pages     = {012001},
  volume    = {5},
  doi       = {10.1088/2050-6120/aa537e},
  publisher = {IOP Publishing},
}

@Book{Krishnamoorthy2006,
  author    = {Krishnamoorthy, K.},
  publisher = {Chapman \& Hall/CRC,},
  title     = {Handbook of Statistical Distributions with Applications},
  year      = {2006},
  %address   = {Boca Raton},
  isbn      = {1-5848-8635-8},
  number    = {188},
  pages     = {185--194},
  series    = {Statistics, a series of textbooks \& monographs},
  %pagetotal = {1346},
  ppn_gvk   = {1932514481},
}

@Article{Olivier2017,
  author    = {Olivier, Y. and Yurash, B. and Muccioli, L. and D’Avino, G. and Mikhnenko, O. and Sancho-García, J. C. and Adachi, C. and Nguyen, T.-Q. and Beljonne, D.},
  journal   = {Physical Review Materials},
  title     = {Nature of the singlet and triplet excitations mediating thermally activated delayed fluorescence},
  year      = {2017},
  issn      = {2475-9953},
  month     = dec,
  number    = {7},
  pages     = {075602},
  volume    = {1},
  doi       = {10.1103/physrevmaterials.1.075602},
  publisher = {American Physical Society (APS)},
}

@Article{Hosokai2018,
  author    = {Hosokai, Takuya and Noda, Hiroki and Nakanotani, Hajime and Nawata, Takanori and Nakayama, Yasuo and Matsuzaki, Hiroyuki and Adachi, Chihaya},
  journal   = {Journal of Photonics for Energy},
  title     = {Solvent-dependent investigation of carbazole benzonitrile derivatives: does the {LE3--CT1} energy gap facilitate thermally activated delayed fluorescence?},
  year      = {2018},
  issn      = {1947-7988},
  month     = feb,
  number    = {03},
  pages     = {1},
  volume    = {8},
  doi       = {10.1117/1.jpe.8.032102},
  publisher = {SPIE-Intl Soc Optical Eng},
}

@Article{Huang2024,
  author    = {Huang, Tianyu and Wang, Qi and Zhang, Hai and Zhang, Yuewei and Zhan, Ge and Zhang, Dongdong and Duan, Lian},
  journal   = {Nature Photonics},
  title     = {Enhancing the efficiency and stability of blue thermally activated delayed fluorescence emitters by perdeuteration},
  year      = {2024},
  issn      = {1749-4893},
  month     = jan,
  number    = {5},
  pages     = {516--523},
  volume    = {18},
  doi       = {10.1038/s41566-024-01379-1},
  publisher = {Springer Science and Business Media LLC},
}

@Article{Cho2020,
  author    = {Cho, Eunkyung and Liu, Lei and Coropceanu, Veaceslav and Brédas, Jean-Luc},
  journal   = {The Journal of Chemical Physics},
  title     = {Impact of secondary donor units on the excited-state properties and thermally activated delayed fluorescence (TADF) efficiency of pentacarbazole-benzonitrile emitters},
  year      = {2020},
  issn      = {1089-7690},
  month     = oct,
  number    = {14},
  volume    = {153},
  pages     = {144708},
  doi       = {10.1063/5.0028227},
  publisher = {AIP Publishing},
}

@Article{Speckmeier2018,
  author    = {Speckmeier, Elisabeth and Fischer, Tillmann G. and Zeitler, Kirsten},
  journal   = {Journal of the American Chemical Society},
  title     = {A Toolbox Approach To Construct Broadly Applicable Metal-Free Catalysts for Photoredox Chemistry: Deliberate Tuning of Redox Potentials and Importance of Halogens in Donor–Acceptor Cyanoarenes},
  year      = {2018},
  issn      = {1520-5126},
  month     = oct,
  number    = {45},
  pages     = {15353--15365},
  volume    = {140},
  doi       = {10.1021/jacs.8b08933},
  publisher = {American Chemical Society (ACS)},
}

@misc{gv6,
author={Dennington, R. and Keith, TA. and Millam, JM. },
title={GaussView {V}ersion {6}},
note={Semichem Inc. Shawnee Mission KS},
year={2019}
}

@misc{g16,
author={M. J. Frisch and G. W. Trucks and H. B. Schlegel and G. E. Scuseria and M. A. Robb and J. R. Cheeseman and G. Scalmani and V. Barone and G. A. Petersson and H. Nakatsuji and X. Li and M. Caricato and A. V. Marenich and J. Bloino and B. G. Janesko and R. Gomperts and B. Mennucci and H. P. Hratchian and J. V. Ortiz and A. F. Izmaylov and J. L. Sonnenberg and D. Williams-Young and F. Ding and F. Lipparini and F. Egidi and J. Goings and B. Peng and A. Petrone and T. Henderson and D. Ranasinghe and V. G. Zakrzewski and J. Gao and N. Rega and G. Zheng and W. Liang and M. Hada and M. Ehara and K. Toyota and R. Fukuda and J. Hasegawa and M. Ishida and T. Nakajima and Y. Honda and O. Kitao and H. Nakai and T. Vreven and K. Throssell and Montgomery, {Jr.}, J. A. and J. E. Peralta and F. Ogliaro and M. J. Bearpark and J. J. Heyd and E. N. Brothers and K. N. Kudin and V. N. Staroverov and T. A. Keith and R. Kobayashi and J. Normand and K. Raghavachari and A. P. Rendell and J. C. Burant and S. S. Iyengar and J. Tomasi and M. Cossi and J. M. Millam and M. Klene and C. Adamo and R. Cammi and J. W. Ochterski and R. L. Martin and K. Morokuma and O. Farkas and J. B. Foresman and D. J. Fox},
title={Gaussian 16 {R}evision {C}.02},
year={2016},
note={Gaussian Inc. Wallingford CT}
}

@article{vosko1980accurate,
  title={Accurate spin-dependent electron liquid correlation energies for local spin density calculations: a critical analysis},
  author={Vosko, SH. and Wilk, L. and Nusair, M.},
  journal={Canadian Journal of Physics},
  volume={58},
  number={8},
  pages={1200--1211},
  year={1980},
  publisher={NRC Research Press Ottawa, Canada},
  doi={10.1139/p80-159}
}

@article{lee1988development,
  title={Development of the Colle-Salvetti correlation-energy formula into a functional of the electron density},
  author={Lee, C. and Yang, W. and Parr, RG.},
  journal={Physical Review B},
  volume={37},
  number={2},
  pages={785},
  year={1988},
  publisher={APS},
  doi={10.1103/PhysRevB.37.785}
}

@article{becke1988density,
  title={Density-functional exchange-energy approximation with correct asymptotic behavior},
  author={Becke, AD.},
  journal={Physical Review A},
  volume={38},
  number={6},
  pages={3098},
  year={1988},
  publisher={APS},
  doi={10.1103/PhysRevA.38.3098}
}

@article{becke1993density,
  title={Density-functional thermochemistry. III. The role of exact exchange},
  author={Becke, AD.},
  journal={The Journal of Chemical Physics},
  volume={98},
  number={7},
  pages={5648--5652},
  year={1993},
  publisher={American Institute of Physics},
  doi={10.1063/1.464913}
}

@article{stephens1994ab,
  title={Ab initio calculation of vibrational absorption and circular dichroism spectra using density functional force fields},
  author={Stephens, PJ. and Devlin, FJ. and Chabalowski, CF. and Frisch, MJ.},
  journal={The Journal of Physical Chemistry},
  volume={98},
  number={45},
  pages={11623--11627},
  year={1994},
  publisher={ACS Publications},
  doi={10.1021/j100096a001}
}

@article{grimme2010consistent,
  title={A consistent and accurate ab initio parametrization of density functional dispersion correction (DFT-D) for the 94 elements H-Pu},
  author={Grimme, S. and Antony, J. and Ehrlich, S. and Krieg, H.},
  journal={The Journal of Chemical Physics},
  volume={132},
  pages = {154104},
  number={15},
  year={2010},
  publisher={AIP Publishing},
  doi={10.1063/1.3382344}
}

@article{clark1983a,
    author = {Clark, T. and Chandrasekhar, J. and Spitznagel, G.~W. and Schleyer, P.~V.~R.},
    title = {Efficient diffuse function-augmented basis sets for anion calculations. III. The 3-21+G basis set for first-row elements, Li-F},
    journal = {Journal of Computational Chemistry},
    volume = {4},
    pages = {294-301},
    year = {1983},
    doi = {10.1002/jcc.540040303}
}

@article{ditchfield1971a,
    author = {Ditchfield, R. and Hehre, W.~J. and Pople, J.~A.},
    title = {Self-Consistent Molecular-Orbital Methods. IX. An Extended Gaussian-Type Basis for Molecular-Orbital Studies of Organic Molecules},
    journal = {Journal of Chemical Physics},
    volume = {54},
    pages = {724-728},
    year = {1971},
    doi = {10.1063/1.1674902}
}

@article{francl1982a,
    author = {Francl, M.~M. and Pietro, W.~J. and Hehre, W.~J. and Binkley, J.~S. and Gordon, M.~S. and DeFrees, D.~J. and Pople, J.~A.},
    title = {Self-consistent molecular orbital methods. XXIII. A polarization-type basis set for second-row elements},
    journal = {Journal of Chemical Physics},
    volume = {77},
    pages = {3654-3665},
    year = {1982},
    doi = {10.1063/1.444267}
}

@article{gordon1982a,
    author = {Gordon, M.~S. and Binkley, J.~S. and Pople, J.~A. and Pietro, W.~J. and Hehre, W.~J.},
    title = {Self-consistent molecular-orbital methods. 22. Small split-valence basis sets for second-row elements},
    journal = {Journal of the American Chemical Society},
    volume = {104},
    pages = {2797-2803},
    year = {1982},
    doi = {10.1021/ja00374a017}
}

@article{hariharan1973a,
    author = {Hariharan, PC. and Pople, JA.},
    title = {The influence of polarization functions on molecular orbital hydrogenation energies},
    journal = {Thermochimica Acta},
    volume = {28},
    pages = {213-222},
    year = {1973},
    doi = {10.1007/BF00533485}
}

@article{hehre1972a,
    author = {Hehre, W. and Ditchfield, R. and Pople, JA.},
    title = {Self-Consistent Molecular Orbital Methods. XII. Further Extensions of Gaussian-Type Basis Sets for Use in Molecular Orbital Studies of Organic Molecules},
    journal = {Journal of Chemical Physics},
    volume = {56},
    pages = {2257-2261},
    year = {1972},
    doi = {10.1063/1.1677527}
}

@article{spitznagel1987a,
    author = {Spitznagel, G.~W. and Clark, T. and Schleyer, P.~V~.R. and Hehre, W.~J.},
    title = {An evaluation of the performance of diffuse function-augmented basis sets for second row elements, Na-Cl},
    journal = {Journal of Computational Chemistry},
    volume = {8},
    pages = {1109-1116},
    year = {1987},
    doi = {10.1002/jcc.540080807}
}

@article{morgenstern2025unlocking,
  title={Unlocking Spin Dynamics: Spin-Orbit Coupling Driven Spin State Interconversion in Carbazole-Containing TADF Emitters},
  author = {Morgenstern, Annika and Weiser, Jonas and Schreier, Lucas and Gabel, Konstantin and Gabler, Tom and Ehm, Alexander and Beer, Daniel and Schwierz, Nadine and Schwarz, Ulrich T. and Zeitler, Kirsten and Deibel, Carsten and Zahn, Dietrich R. T. and Wiebeler, Christian and Salvan, Georgeta},
  journal={Advanced Optical Materials},
  volume={14},
  number={2},
  pages={e01504},
  year={2025},
  publisher={Wiley Online Library},
  doi={10.1002/adom.202501504}
}

@article{de2025goat,
  title={GOAT: A Global Optimization Algorithm for Molecules and Atomic Clusters},
  author={de Souza, B.},
  journal={Angewandte Chemie International Edition},
  volume = {64},
  number = {18},
  pages = {e202500393},
  year={2025},
  publisher={Wiley Online Library},
  doi={10.1002/anie.202500393}
}

@article{ORCA,
author = {Neese,F.},
title = {The ORCA program system},
journal = {WIREs Computational Molecular Science},
volume = {2},
number = {1},
pages = {73-78},
year = {2012},
type = {journal Article},
doi = {10.1002/wcms.81}
}

@article{neese2025software,
  title={Software update: The ORCA program system—version 6.0},
  author={Neese, F.},
  journal={Wiley Interdisciplinary Reviews: Computational Molecular Science},
  volume={15},
  number={2},
  pages={e70019},
  year={2025},
  publisher={Wiley Online Library},
  doi={10.1002/wcms.70019}
}

@article{bannwarth2019gfn2,
  title={GFN2-xTB—An accurate and broadly parametrized self-consistent tight-binding quantum chemical method with multipole electrostatics and density-dependent dispersion contributions},
  author={Bannwarth, C. and Ehlert, S. and Grimme, S.},
  journal={Journal of Chemical Theory and Computation},
  volume={15},
  number={3},
  pages={1652--1671},
  year={2019},
  publisher={ACS Publications},
  doi={10.1021/acs.jctc.8b01176}
}

@article{rappoport2010a,
    author = {Rappoport, D. and Furche, F.},
    title = {Property-optimized Gaussian basis sets for molecular response calculations},
    journal = {The Journal of Chemical Physics},
    volume = {133},
    pages = {134105},
    year = {2010},
    doi = {10.1063/1.3484283}
}

@article{weigend2005a,
    author = {Weigend, F. and Ahlrichs, R.},
    title = {Balanced basis sets of split valence, triple zeta valence and quadruple zeta valence quality for H to Rn: Design and assessment of accuracy},
    journal = {Physical Chemistry Chemical Physics},
    volume = {7},
    pages = {3297},
    year = {2005},
    doi = {10.1039/B508541A}
}

@article{marcus1993electron,
  title={Electron transfer reactions in chemistry: theory and experiment},
  author={Marcus, Rudolph A},
  journal={Angewandte Chemie International Edition in English},
  volume={32},
  number={8},
  pages={1111--1121},
  year={1993},
  publisher={Wiley Online Library},
  doi={10.1002/anie.199311113}
}

@article{bredas2004charge,
  title={Charge-transfer and energy-transfer processes in $\pi$-conjugated oligomers and polymers: a molecular picture},
  author={Br{\'e}das, Jean-Luc and Beljonne, David and Coropceanu, Veaceslav and Cornil, J{\'e}r{\^o}me},
  journal={Chemical Reviews},
  volume={104},
  number={11},
  pages={4971--5004},
  year={2004},
  publisher={ACS Publications},
  doi={10.1021/cr040084k}
}

@article{koopmans1934zuordnung,
  title={{\"U}ber die Zuordnung von Wellenfunktionen und Eigenwerten zu den einzelnen Elektronen eines Atoms},
  author={Koopmans, Tjalling},
  journal={Physica},
  volume={1},
  number={1-6},
  pages={104--113},
  year={1934},
  publisher={Elsevier},
  doi={10.1016/S0031-8914(34)90011-2}
}

@article{stein2009reliable,
  title={Reliable prediction of charge transfer excitations in molecular complexes using time-dependent density functional theory},
  author={Stein, Tamar and Kronik, Leeor and Baer, Roi},
  journal={Journal of the American Chemical Society},
  volume={131},
  number={8},
  pages={2818--2820},
  year={2009},
  publisher={ACS Publications},
  doi={10.1021/ja8087482}
}

@article{brent1971algorithm,
  title={An algorithm with guaranteed convergence for finding a zero of a function},
  author={Brent, Richard P.},
  journal={The Computer Journal},
  volume={14},
  number={4},
  pages={422--425},
  year={1971},
  publisher={Oxford University Press},
  doi={10.1093/comjnl/14.4.422}
}

@article{shannon1948mathematical,
  title={A Mathematical Theory of Communication},
  author={Shannon, Claude E.},
  journal={The Bell System Technical Journal},
  volume={27},
  pages={379},
  year={1948},
  doi={10.1002/j.1538-7305.1948.tb01338.x}
}

@article{spellerberg2003tribute,
  title={A Tribute to Claude Shannon (1916--2001) and a Plea for More Rigorous Use of Species Richness, Species Diversity and the ‘Shannon--Wiener’Index},
  author={Spellerberg, Ian F. and Fedor, Peter J.},
  journal={Global Ecology and Biogeography},
  volume={12},
  pages={177},
  year={2003},
  doi={10.1046/j.1466-822X.2003.00015.x}
}

@article{lindstrom1988newton,
  title={Newton—Raphson and EM algorithms for linear mixed-effects models for repeated-measures data},
  author={Lindstrom, Mary J and Bates, Douglas M},
  journal={Journal of the American Statistical Association},
  volume={83},
  number={404},
  pages={1014--1022},
  year={1988},
  publisher={Taylor \& Francis},
  doi={10.1080/01621459.1988.10478693}
}

@article{box1964analysis,
  title={An analysis of transformations},
  author={Box, George E. P. and Cox, David R.},
  journal={Journal of the Royal Statistical Society Series B: Statistical Methodology},
  volume={26},
  pages={211},
  year={1964},
  doi={10.1111/j.2517-6161.1964.tb00553.x}
}

@article{kaminski2024balancing,
  title={Balancing {TADF} Properties in $\pi$-Bridged Donor--Acceptor Systems by Sterical Constraints: The Best of Three Worlds},
  author={Kaminski, Jeremy and Böhmer, Tobias and Marian, Christel},
  journal={The Journal of Physical Chemistry C},
  volume={128},
  number={33},
  pages={13711--13721},
  year={2024},
  publisher={ACS Publications},
  doi={10.1021/acs.jpcc.4c03865}
}

@article{de2024tadf,
  title={Tadf mechanism in a carbene-copper emitter: Insights from the nuclear ensemble simulations},
  author={de Thieulloy, Laure and de Sousa, Leonardo Evaristo and de Silva, Piotr},
  journal={The Journal of Physical Chemistry C},
  volume={128},
  number={36},
  pages={14887--14896},
  year={2024},
  publisher={ACS Publications},
  doi={10.1021/acs.jpcc.4c03860}
}

@Article{Tanimoto_2016,
  author    = {Tanimoto, Shuho and Suzuki, Takatsugu and Nakanotani, Hajime and Adachi, Chihaya},
  journal   = {Chemistry Letters},
  title     = {Thermally Activated Delayed Fluorescence from Pentacarbazorylbenzonitrile},
  year      = {2016},
  issn      = {1348-0715},
  month     = jul,
  number    = {7},
  pages     = {770--772},
  volume    = {45},
  comment   = {5CzBN},
  doi       = {10.1246/cl.160290},
  publisher = {Oxford University Press (OUP)},
}

@Article{Serevicius2023,
  author    = {Serevičius, Tomas and Skaisgiris, Rokas and Tumkevičius, Sigitas and Dodonova-Vaitkūnienė, Jelena and Juršėnas, Saulius},
  journal   = {Journal of Materials Chemistry C},
  title     = {Understanding the temporal dynamics of thermally activated delayed fluorescence in solid hosts},
  year      = {2023},
  issn      = {2050-7534},
  number    = {36},
  pages     = {12147--12155},
  volume    = {11},
  doi       = {10.1039/d3tc02347h},
  publisher = {Royal Society of Chemistry (RSC)},
}

@Article{Thieulloy2025,
  author    = {de Thieulloy, Laure and de Sousa, Leonardo Evaristo and de Silva, Piotr},
  journal   = {Journal of Materials Chemistry C},
  title     = {Enhancing triplet harvesting in inverted singlet–triplet gap molecules through mechanistic understanding},
  year      = {2025},
  issn      = {2050-7534},
  number    = {32},
  pages     = {16489--16498},
  volume    = {13},
  doi       = {10.1039/d5tc01155h},
  publisher = {Royal Society of Chemistry (RSC)},
}

@article{plasser2020theodore,
  title     = {TheoDORE: A Toolbox for a Detailed and Automated Analysis of Electronic Excited State Computations},
  author    = {Plasser, Felix},
  journal   = {The Journal of Chemical Physics},
  volume    = {152},
  number    = {8},
  year      = {2020},
  month     = feb,
  issn      = {1089-7690},
  pages     = {084108},
  doi       = {10.1063/1.5143076},
  publisher = {AIP Publishing}
}

@article{lin2013long,
  title     = {Long-Range Corrected Hybrid Density Functionals With Improved Dispersion Corrections},
  author    = {Lin, You-Sheng and Li, Guan-De and Mao, Shan-Ping and Chai, Jeng-Da},
  journal   = {Journal of Chemical Theory and Computation},
  issn      = {1549-9626},
  month     = nov,
  number    = {1},
  pages     = {263--272},
  volume    = {9},
  doi       = {10.1021/ct300715s},
  publisher = {American Chemical Society (ACS)},
  year      = {2013}
}

@Article{Qiu2023,
  author    = {Qiu, Weidong and Liu, Denghui and Li, Mengke and Cai, Xinyi and Chen, Zijian and He, Yanmei and Liang, Baoyan and Peng, Xiaomei and Qiao, Zhenyang and Chen, Jiting and Li, Wei and Pu, Junrong and Xie, Wentao and Wang, Zhiheng and Li, Deli and Gan, Yiyang and Jiao, Yihang and Gu, Qing and Su, Shi-Jian},
  journal   = {Nature Communications},
  title     = {Confining donor conformation distributions for efficient thermally activated delayed fluorescence with fast spin-flipping},
  year      = {2023},
  issn      = {2041-1723},
  month     = may,
  number    = {1},
  pages     = {2564},
  volume    = {14},
  doi       = {10.1038/s41467-023-38197-y},
  publisher = {Springer Science and Business Media LLC},
}

@article{souza2025dynamics,
  title     = {Dynamics of Vibrationally Coupled Intersystem Crossing in State-Of-The-Art Organic Optoelectronic Materials},
  author    = {Souza, J.P.A. and Benatto, L. and Candiotto, G. and Wouk, L. and Koehler, M.},
  journal   = {Communications Chemistry},
  volume    = {8},
  number    = {1},
  pages     = {84},
  year      = {2025},
  doi       = {10.1038/s42004-025-01485-3},
  publisher = {Nature Publishing Group UK London}
}

@article{gao2026unveiling,
  author    = {Gao, Yang and Jiang, Congzhou and Zhao, Chen and Lin, Lili and Wang, Chuan-Kui and Fan, Jianzhong and Song, Yuzhi},
  journal   = {The Journal of Physical Chemistry Letters},
  title     = {Unveiling Multichannel Triplet Excitons Up-Conversion Mechanisms in Blue Thermally Activated Delayed Fluorescence Emitters},
  year      = {2026},
  issn      = {1948-7185},
  month     = jan,
  number    = {3},
  pages     = {780--789},
  volume    = {17},
  doi       = {10.1021/acs.jpclett.5c03963},
  publisher = {American Chemical Society (ACS)},
}

@article{gibson2017nonadiabatic,
  title     = {Nonadiabatic Coupling Reduces the Activation Energy in Thermally Activated Delayed Fluorescence},
  author    = {Gibson, J and Penfold, Thomas J},
  journal   = {Physical Chemistry Chemical Physics},
  issn      = {1463-9084},
  volume    = {19},
  number    = {12},
  pages     = {8428--8434},
  year      = {2017},
  doi       = {10.1039/c7cp00719a},
  publisher = {Royal Society of Chemistry}
}

@article{kim2019spin,
  title     = {Spin--Vibronic Model for Quantitative Prediction of Reverse Intersystem Crossing Rate in Thermally Activated Delayed Fluorescence Systems},
  author    = {Kim, Inkoo and Jeon, Soon Ok and Jeong, Daun and Choi, Hyeonho and Son, Won-Joon and Kim, Dongwook and Rhee, Young Min and Lee, Hyo Sug},
  journal   = {Journal of Chemical Theory and Computation},
  volume    = {16},
  number    = {1},
  pages     = {621--632},
  year      = {2019},
  issn      = {1549-9626},
  doi       = {10.1021/acs.jctc.9b01014},
  publisher = {ACS Publications}
}

@Article{Chernowsky2021,
  author    = {Chernowsky, Colleen P. and Chmiel, Alyah F. and Wickens, Zachary K.},
  journal   = {Angewandte Chemie International Edition},
  title     = {Electrochemical Activation of Diverse Conventional Photoredox Catalysts Induces Potent Photoreductant Activity**},
  year      = {2021},
  issn      = {1521-3773},
  month     = aug,
  number    = {39},
  pages     = {21418--21425},
  volume    = {60},
  doi       = {10.1002/anie.202107169},
  publisher = {Wiley},
}

@Article{Garreau2019,
  author    = {Garreau, Marion and Le Vaillant, Franck and Waser, Jerome},
  journal   = {Angewandte Chemie International Edition},
  title     = {C‐Terminal Bioconjugation of Peptides through Photoredox Catalyzed Decarboxylative Alkynylation},
  year      = {2019},
  issn      = {1521-3773},
  month     = may,
  number    = {24},
  pages     = {8182--8186},
  volume    = {58},
  doi       = {10.1002/anie.201901922},
  publisher = {Wiley},
}

@Article{Fulmer2010,
  author    = {Fulmer, Gregory R. and Miller, Alexander J. M. and Sherden, Nathaniel H. and Gottlieb, Hugo E. and Nudelman, Abraham and Stoltz, Brian M. and Bercaw, John E. and Goldberg, Karen I.},
  journal   = {Organometallics},
  title     = {NMR Chemical Shifts of Trace Impurities: Common Laboratory Solvents, Organics, and Gases in Deuterated Solvents Relevant to the Organometallic Chemist},
  year      = {2010},
  issn      = {1520-6041},
  month     = apr,
  number    = {9},
  pages     = {2176--2179},
  volume    = {29},
  doi       = {10.1021/om100106e},
  publisher = {American Chemical Society (ACS)},
}

@Article{DosSantos2024,
  author    = {Dos Santos, John Marques and Hall, David and Basumatary, Biju and Bryden, Megan and Chen, Dongyang and Choudhary, Praveen and Comerford, Thomas and Crovini, Ettore and Danos, Andrew and De, Joydip and Diesing, Stefan and Fatahi, Mahni and Griffin, Máire and Gupta, Abhishek Kumar and Hafeez, Hassan and Hämmerling, Lea and Hanover, Emily and Haug, Janine and Heil, Tabea and Karthik, Durai and Kumar, Shiv and Lee, Oliver and Li, Haoyang and Lucas, Fabien and Mackenzie, Campbell Frank Ross and Mariko, Aminata and Matulaitis, Tomas and Millward, Francis and Olivier, Yoann and Qi, Quan and Samuel, Ifor D. W. and Sharma, Nidhi and Si, Changfeng and Spierling, Leander and Sudhakar, Pagidi and Sun, Dianming and Tankelevi\v{c}iu\={u}t\.{e}, Eglė and Duarte Tonet, Michele and Wang, Jingxiang and Wang, Tao and Wu, Sen and Xu, Yan and Zhang, Le and Zysman-Colman, Eli},
  journal   = {Chemical Reviews},
  title     = {The Golden Age of Thermally Activated Delayed Fluorescence Materials: Design and Exploitation},
  year      = {2024},
  issn      = {1520-6890},
  month     = dec,
  number    = {24},
  pages     = {13736--14110},
  volume    = {124},
  doi       = {10.1021/acs.chemrev.3c00755},
  publisher = {American Chemical Society (ACS)},
}

@article{Goodfellow2026,
  author    = {Goodfellow, Alister S. and Nguyen, Bao N.},
  journal   = {Journal of Chemical Theory and Computation},
  title     = {Graph-Based Internal Coordinate Analysis for Transition State Characterization},
  year      = {2026},
  issn      = {1549-9626},
  month     = feb,
  number    = {5},
  pages     = {2348--2357},
  volume    = {22},
  doi       = {10.1021/acs.jctc.5c02073},
  publisher = {American Chemical Society (ACS)},
}

\end{refsection}

\end{document}